\documentclass{nature}
\pdfoutput=1
\usepackage[breaklinks]{hyperref}
\hypersetup{colorlinks,citecolor=black,filecolor=black,linkcolor=black,urlcolor=black}
\usepackage{url}
\usepackage{xcolor}
\usepackage{siunitx}
\usepackage{booktabs}
\usepackage{aas_macros}
\usepackage{amssymb}
\usepackage{amsmath}
\usepackage{amsthm}
\usepackage{longtable}

\newcommand{\JHK}{$JHK_{\mathrm{s}}$}

\newcommand{\arcsec}{$^{\prime\prime}$}
\newcommand{\ee}{$\epsilon_{\rm e}$}
\newcommand{\eb}{$\epsilon_{\rm B}$}
\newcommand{\nui}{$\nu_{\rm m}$}
\newcommand{\nuc}{$\nu_{\rm c}$}
\newcommand{\nua}{$\nu_{\rm sa}$}
\newcommand{\pp}{$\gamma$-$\gamma$}
\newcommand{\gm}{$\gamma_{\rm m}$}

\newcommand{\lae}{\lower 2pt \hbox{$\, \buildrel {\scriptstyle <}\over {\scriptstyle
\sim}\,$}}
\title{Observation of inverse Compton emission from a long \\  $\gamma$-ray burst}

\begin{document}
\maketitle

\begin{centering}
\author{
V.\,A.\,Acciari$^{\ref{magic:1}}$,
S.\,Ansoldi$^{\ref{magic:2},\ref{magic:31}}$,
L.\,A.\,Antonelli$^{\ref{magic:3}}$,
A.\,Arbet Engels$^{\ref{magic:4}}$,
D.\,Baack$^{\ref{magic:5}}$,
A.\,Babi\'c$^{\ref{magic:36}}$,
B.\,Banerjee$^{\ref{magic:7}}$,
U.\,Barres de Almeida$^{\ref{magic:8}}$,
J.\,A.\,Barrio$^{\ref{magic:9}}$,
J.\,Becerra Gonz\'alez$^{\ref{magic:1}}$,
W.\,Bednarek$^{\ref{magic:10}}$,
L.\,Bellizzi$^{\ref{magic:11}}$,
E.\,Bernardini$^{\ref{magic:12},\ref{magic:16}}$,
A.\,Berti$^{\ref{magic:13}}$,
J.\,Besenrieder$^{\ref{magic:14}}$,
W.\,Bhattacharyya$^{\ref{magic:12}}$,
C.\,Bigongiari$^{\ref{magic:3}}$,
A.\,Biland$^{\ref{magic:4}}$,
O.\,Blanch$^{\ref{magic:15}}$,
G.\,Bonnoli$^{\ref{magic:11}}$,
\v{Z}.\,Bo\v{s}njak$^{\ref{magic:36}}$,
G.\,Busetto$^{\ref{magic:16}}$,
R.\,Carosi$^{\ref{magic:17}}$,
G.\,Ceribella$^{\ref{magic:14}}$,
Y.\,Chai$^{\ref{magic:14}}$,
A.\,Chilingaryan$^{\ref{magic:40}}$,
S.\,Cikota$^{\ref{magic:36}}$,
S.\,M.\,Colak$^{\ref{magic:15}}$,
U.\,Colin$^{\ref{magic:14}}$,
E.\,Colombo$^{\ref{magic:1}}$,
J.\,L.\,Contreras$^{\ref{magic:9}}$,
J.\,Cortina$^{\ref{magic:ciemat}}$,
S.\,Covino$^{\ref{magic:3}}$,
V.\,D'Elia$^{\ref{magic:3}}$,
P.\,Da Vela$^{\ref{magic:17}}$,
F.\,Dazzi$^{\ref{magic:3}}$,
A.\,De Angelis$^{\ref{magic:16}}$,
B.\,De Lotto$^{\ref{magic:2}}$,
M.\,Delfino$^{\ref{magic:15},\ref{magic:26}}$,
J.\,Delgado$^{\ref{magic:15},\ref{magic:26}}$,
D.\,Depaoli$^{\ref{magic:13}}$,
F.\,Di Pierro$^{\ref{magic:13}}$,
L.\,Di Venere$^{\ref{magic:13}}$,
E.\,Do Souto Espi\~neira$^{\ref{magic:15}}$,
D.\,Dominis Prester$^{\ref{magic:34}}$,
A.\,Donini$^{\ref{magic:2}}$,
D.\,Dorner$^{\ref{magic:18}}$,
M.\,Doro$^{\ref{magic:16}}$,
D.\,Elsaesser$^{\ref{magic:5}}$,
V.\,Fallah Ramazani$^{\ref{magic:41}}$,
A.\,Fattorini$^{\ref{magic:5}}$,
G.\,Ferrara$^{\ref{magic:3}}$,
D.\,Fidalgo$^{\ref{magic:9}}$,
L.\,Foffano$^{\ref{magic:16}}$,
M.\,V.\,Fonseca$^{\ref{magic:9}}$,
L.\,Font$^{\ref{magic:20}}$,
C.\,Fruck$^{\ref{magic:14}}$,
S.\,Fukami$^{\ref{magic:30}}$,
R.\,J.\,Garc\'ia L\'opez$^{\ref{magic:1}}$,
M.\,Garczarczyk$^{\ref{magic:12}}$,
S.\,Gasparyan$^{\ref{magic:39}}$,
M.\,Gaug$^{\ref{magic:20}}$,
N.\,Giglietto$^{\ref{magic:13}}$,
F.\,Giordano$^{\ref{magic:13}}$,
N.\,Godinovi\'c$^{\ref{magic:35}}$,
D.\,Green$^{\ref{magic:14}}$,
D.\,Guberman$^{\ref{magic:15}}$,
D.\,Hadasch$^{\ref{magic:30}}$,
A.\,Hahn$^{\ref{magic:14}}$,
J.\,Herrera$^{\ref{magic:1}}$,
J.\,Hoang$^{\ref{magic:9}}$,
D.\,Hrupec$^{\ref{magic:37}}$,
M.\,H\"utten$^{\ref{magic:14}}$,
T.\,Inada$^{\ref{magic:30}}$,
S.\,Inoue$^{\ref{magic:33}}$,
K.\,Ishio$^{\ref{magic:14}}$,
Y.\,Iwamura$^{\ref{magic:30}}$,
L.\,Jouvin$^{\ref{magic:15}}$,
D.\,Kerszberg$^{\ref{magic:15}}$,
H.\,Kubo$^{\ref{magic:31}}$,
J.\,Kushida$^{\ref{magic:32}}$,
A.\,Lamastra$^{\ref{magic:3}}$,
D.\,Lelas$^{\ref{magic:35}}$,
F.\,Leone$^{\ref{magic:3}}$,
E.\,Lindfors$^{\ref{magic:41}}$,
S.\,Lombardi$^{\ref{magic:3}}$,
F.\,Longo$^{\ref{magic:2},\ref{magic:28}, \ref{magic:25}}$,
M.\,L\'opez$^{\ref{magic:9}}$,
R.\,L\'opez-Coto$^{\ref{magic:16}}$,
A.\,L\'opez-Oramas$^{\ref{magic:1}}$,
S.~Loporchio$^{\ref{magic:13}}$, 
B.~Machado de Oliveira Fraga$^{\ref{magic:8}}$, 
C.\,Maggio$^{\ref{magic:20}}$,
P.\,Majumdar$^{\ref{magic:7}}$,
M.\,Makariev$^{\ref{magic:23}}$,
M.\,Mallamaci$^{\ref{magic:16}}$,
G.\,Maneva$^{\ref{magic:23}}$,
M.\,Manganaro$^{\ref{magic:34}}$,
K.\,Mannheim$^{\ref{magic:18}}$,
L.\,Maraschi$^{\ref{magic:3}}$,
M.\,Mariotti$^{\ref{magic:16}}$,
M.\,Mart\'inez$^{\ref{magic:15}}$,
D.\,Mazin$^{\ref{magic:14},\ref{magic:30}}$,
S.\,Mi\'canovi\'c$^{\ref{magic:34}}$,
D.\,Miceli$^{\ref{magic:2}}$,
M.\,Minev$^{\ref{magic:23}}$,
J.\,M.\,Miranda$^{\ref{magic:11}}$,
R.\,Mirzoyan$^{\ref{magic:14}}$,
E.\,Molina$^{\ref{magic:24}}$,
A.\,Moralejo$^{\ref{magic:15}}$,
D.\,Morcuende$^{\ref{magic:9}}$,
V.\,Moreno$^{\ref{magic:20}}$,
E.\,Moretti$^{\ref{magic:15}}$,
P.\,Munar-Adrover$^{\ref{magic:20}}$,
V.\,Neustroev$^{\ref{magic:42}}$,
C.\,Nigro$^{\ref{magic:12}}$,
K.\,Nilsson$^{\ref{magic:41}}$,
D.\,Ninci$^{\ref{magic:15}}$,
K.\,Nishijima$^{\ref{magic:32}}$,
K.\,Noda$^{\ref{magic:30}}$,
L.\,Nogu\'es$^{\ref{magic:15}}$,
S.\,Nozaki$^{\ref{magic:31}}$,
S.\,Paiano$^{\ref{magic:16}}$,
M.\,Palatiello$^{\ref{magic:2}}$,
D.\,Paneque$^{\ref{magic:14}}$,
R.\,Paoletti$^{\ref{magic:11}}$,
J.\,M.\,Paredes$^{\ref{magic:24}}$,
P.\,Pe\~nil$^{\ref{magic:9}}$,
M.\,Peresano$^{\ref{magic:2}}$,
M.\,Persic$^{\ref{magic:2}}$,
P.\,G.\,Prada Moroni$^{\ref{magic:17}}$,
E.\,Prandini$^{\ref{magic:16}}$,
I.\,Puljak$^{\ref{magic:35}}$,
W.\,Rhode$^{\ref{magic:5}}$,
M.\,Rib\'o$^{\ref{magic:24}}$,
J.\,Rico$^{\ref{magic:15}}$,
C.\,Righi$^{\ref{magic:3}}$,
A.\,Rugliancich$^{\ref{magic:17}}$,
L.\,Saha$^{\ref{magic:9}}$,
N.\,Sahakyan$^{\ref{magic:39}}$,
T.\,Saito$^{\ref{magic:30}}$,
S.\,Sakurai$^{\ref{magic:30}}$,
K.\,Satalecka$^{\ref{magic:12}}$,
K.\,Schmidt$^{\ref{magic:5}}$,
T.\,Schweizer$^{\ref{magic:14}}$,
J.\,Sitarek$^{\ref{magic:10}}$,
I.\,\v{S}nidari\'c$^{\ref{magic:38}}$,
D.\,Sobczynska$^{\ref{magic:10}}$,
A.\,Somero$^{\ref{magic:1}}$,
A.\,Stamerra$^{\ref{magic:3}}$,
D.\,Strom$^{\ref{magic:14}}$,
M.\,Strzys$^{\ref{magic:14}}$,
Y.\,Suda$^{\ref{magic:14}}$,
T.\,Suri\'c$^{\ref{magic:38}}$,
M.\,Takahashi$^{\ref{magic:30}}$,
F.\,Tavecchio$^{\ref{magic:3}}$,
P.\,Temnikov$^{\ref{magic:23}}$,
T.\,Terzi\'c$^{\ref{magic:34},\ref{magic:38}}$,
M.\,Teshima$^{\ref{magic:14},\ref{magic:30}}$,
N.\,Torres-Alb\`a$^{\ref{magic:24}}$,
L.\,Tosti$^{\ref{magic:13}}$,
V.\,Vagelli$^{\ref{magic:13}}$,
J.\,van\,Scherpenberg$^{\ref{magic:14}}$,
G.\,Vanzo$^{\ref{magic:1}}$,
M.\,Vazquez Acosta$^{\ref{magic:1}}$,
C.\,F.\,Vigorito$^{\ref{magic:13}}$,
V.\,Vitale$^{\ref{magic:13}}$,
I.\,Vovk$^{\ref{magic:14}}$,
M.\,Will$^{\ref{magic:14}}$, 
D.\,Zari\'c$^{\ref{magic:35}}$ 

L.\,Nava$^{\ref{magic:3},\ref{magic:28},\ref{magic:29}}$, 

P.\,Veres$^{\ref{fermi-inst:hsv}}$, 
P.\,N.\,Bhat$^{\ref{fermi-inst:hsv}}$,
M.\,S.\,Briggs$^{\ref{fermi-inst:hsv},\ref{fermi-inst:hsv2}}$, 
W.\,H.\,Cleveland$^{\ref{fermi-inst:usra}}$,
R.\,Hamburg$^{\ref{fermi-inst:hsv},\ref{fermi-inst:hsv2}}$,
C.\,M.\,Hui$^{\ref{fermi-inst:msfc}}$, 
B.\,Mailyan$^{\ref{fermi-inst:hsv}}$, 
R.\,D.\,Preece$^{\ref{fermi-inst:hsv},\ref{fermi-inst:hsv2}}$, 
O.\,Roberts$^{\ref{fermi-inst:usra}}$,
A.\,von Kienlin$^{\ref{fermi-inst:mun}}$, 
C.\,A.\,Wilson-Hodge$^{\ref{fermi-inst:msfc}}$,
D.\,Kocevski$^{\ref{fermi-inst:msfc}}$,
M.\,Arimoto$^{\ref{inst:Kanazawa}}$, 
D.\,Tak$^{\ref{alma-inst:umd_phys},\ref{inst:gsfc}}$,
K.\,Asano$^{\ref{fermi-inst:ICRR}}$, 
M.\,Axelsson$^{\ref{fermi-inst:SU},\ref{fermi-inst:KTH}}$, 
G.\,Barbiellini$^{\ref{magic:28}} $, 
E.\,Bissaldi$^{\ref{fermi-inst:UbiBa},\ref{fermi-inst:INFNBa}}$, %
F.\,Fana Dirirsa$^{\ref{fermi-inst:UniJoh}}$,
R.\,Gill$^{\ref{fermi-inst:UniIsrael}}$, 
J.\,Granot$^{\ref{fermi-inst:UniIsrael}}$, 
J.\,McEnery$^{\ref{alma-inst:umd_phys},\ref{inst:gsfc}}$, N.\,Omodei$^{\ref{inst:stanford1},\ref{inst:stanford2}}$,
S.\,Razzaque$^{\ref{fermi-inst:UniJoh}}$, 
F.\,Piron$^{\ref{fermi-inst:LUPM}}$, 
J.\,L.\,Racusin$^{\ref{inst:gsfc}}$, 
D.\,J.\,Thompson$^{\ref{inst:gsfc}}$,

S.\,Campana$^{\ref{rem-ref:oab}}$,
M.\,G.\,Bernardini$^{\ref{rem-ref:oab}}$,
N.\,P.\,M.\,Kuin$^{\ref{uvot-inst:mssl}}$, 
M.\,H.\,Siegel$^{\ref{uvot-inst:psu}}$, 
S.\,Bradley Cenko$^{\ref{inst:gsfc},\ref{cenko2}}$,
P. O’Brien$^{\ref{alma-inst:leicester}}$,
M. Capalbi$^{\ref{palermo}}$,
A. D’A\`i$^{\ref{palermo}}$,
M. De Pasquale$^{\ref{turkey}}$,
J. Gropp$^{\ref{uvot-inst:psu}}$,
N. Klingler$^{\ref{uvot-inst:psu}}$,
J. P. Osborne$^{\ref{alma-inst:leicester}}$,
M. Perri$^{\ref{OAR},\ref{SSDC}}$, 
R. Starling$^{\ref{alma-inst:leicester}}$,
G. Tagliaferri$^{\ref{rem-ref:oab},\ref{palermo}}$,
A. Tohuvavohu$^{\ref{uvot-inst:psu}}$,

A.\,Ursi$^{\ref{iaps_roma}}$,
M.\,Tavani$^{\ref{iaps_roma},\ref{agile2},\ref{agile3}}$,
M.\,Cardillo $^{\ref{iaps_roma}}$,
C.\,Casentini $^{\ref{iaps_roma}}$, 
G.\,Piano$^{\ref{iaps_roma}}$,
Y.\,Evangelista$^{\ref{iaps_roma}}$,
F.\,Verrecchia$^{\ref{OAR}, \ref{SSDC}}$,
C.\,Pittori $^{\ref{OAR}, \ref{SSDC}}$,
F.\,Lucarelli $^{\ref{OAR}, \ref{SSDC}}$,
A.\,Bulgarelli $^{\ref{SSDC}}$,
N.\,Parmiggiani $^{\ref{SSDC}}$,

G.\,E. Anderson$^{\ref{alma-inst:icra}}$,
J.\,P.\,Anderson$^{\ref{alma-inst:eso}}$,
G.\,Bernardi$^{\ref{alma-inst:bologna},\ref{alma-inst:rsa},\ref{alma-inst:ska}}$,
J.\,Bolmer$^{\ref{fermi-inst:mun}}$, 
M.\,D.\,Caballero-Garc\'ia$^{\ref{inst:asu}}$,
I.\,M.\,Carrasco$^{\ref{inst:uma3}}$,
A.\,Castell\'on$^{\ref{inst:uma2}}$,
N.\,Castro Segura$^{\ref{rem-ref:soton}}$,
A.\,J.\,Castro-Tirado$^{\ref{inst:uma},\ref{alma-inst:iaa}}$,
S.\,V.\,Cherukuri$^{\ref{alma-inst:iisst}}$,
A.\,M.\,Cockeram$^{\ref{alma-inst:LivJMU}}$,
P.\,D'Avanzo$^{\ref{rem-ref:oab}}$,
A.\,Di Dato$^{\ref{rem-ref:oasdg},\ref{rem-ref:oac}}$,
R.\,Diretse$^{\ref{inst:UCT}}$,
R.P.\,Fender$^{\ref{inst:Oxford}}$,
E.\,Fern\'andez-Garc\'ia$^{\ref{alma-inst:iaa}}$,
J. P. U. Fynbo$^{\ref{dawn},\ref{bohr}}$,
A.S. Fruchter$^{\ref{baltimore}}$
J.\,Greiner$^{\ref{fermi-inst:mun}}$,
M.\,Gromadzki$^{\ref{rem-ref:warsaw}}$,
K.E. Heintz$^{\ref{iceland}}$
I.\,Heywood$^{\ref{alma-inst:rsa},\ref{inst:Oxford}}$,
A.J.\,van\,der\,Horst$^{\ref{inst:GWU},\ref{inst:APSIS}}$,
Y.-D.\,Hu$^{\ref{alma-inst:iaa},\ref{inst:ugr}}$, 
C.\,Inserra$^{\ref{rem-ref:cardiff}}$, 
L.\,Izzo$^{\ref{alma-inst:iaa},\ref{alma-inst:dark}}$,
V.\,Jaiswal$^{\ref{alma-inst:iisst}}$,
P. Jakobsson$^{\ref{iceland}}$,
J. Japelj$^{\ref{pannekoek}}$,
E.\,Kankare$^{\ref{rem-ref:turku}}$, 
D.\,A.\,Kann$^{\ref{alma-inst:iaa}}$, 
C.\,Kouveliotou$^{\ref{inst:GWU},\ref{inst:APSIS}}$,
S.\,Klose$^{\ref{alma-inst:tls}}$,
A.\,J.\,Levan$^{\ref{nijmegen}}$,
X.\,Y.\,Li$^{\ref{inst:ihsm},\ref{inst:niaot}}$,
S. Lotti$^{\ref{iaps_roma}}$,
K.\,Maguire$^{\ref{rem-ref:trinity}}$,
D. B. Malesani$^{\ref{dawn},\ref{bohr},\ref{alma-inst:dark},\ref{malesani1}}$,
I.\,Manulis$^{\ref{rem-ref:weizmann}}$,
M.\,Marongiu$^{\ref{alma-inst:ferrara}, \ref{icranet}}$,
S.\,Martin$^{\ref{alma-inst:eso},\ref{alma-inst:alma}}$, 
A.\,Melandri$^{\ref{rem-ref:oab}}$,
M.\,Micha{\l}owski$^{\ref{alma-inst:poz}}$, 
J.C.A.\,Miller-Jones$^{\ref{alma-inst:icra}}$,
K.\,Misra$^{\ref{alma-inst:aries},\ref{misra2}}$,
A.\,Moin$^{\ref{alma-inst:uaeu}}$,
K.P.\,Mooley$^{\ref{inst:NRAO},\ref{inst:Caltech}}$,
S.\,Nasri$^{\ref{alma-inst:uaeu}}$,
M.\,Nicholl$^{\ref{rem-ref:edinburgh},\ref{rem-ref:birmingham}}$,
A.\,Noschese$^{\ref{rem-ref:oasdg}}$,
G.\,Novara$^{\ref{xmm-inst:iuss},\ref{iasfmi}}$,
S.\,B.\,Pandey$^{\ref{alma-inst:aries}}$,
E.\,Peretti$^{\ref{agile3},\ref{infn-lngs}}$,
C.\,J.\,P\'erez del Pulgar$^{\ref{inst:uma}}$,
M.A.\,P\'{e}rez-Torres$^{\ref{alma-inst:iaa},\ref{inst:Zaragoza}}$,
D.\,A.\,Perley$^{\ref{alma-inst:LivJMU}}$,
L.\,Piro$^{\ref{iaps_roma}}$,
F.\,Ragosta$^{\ref{rem-ref:oac},\ref{rem-ref:uni-napoli},\ref{rem-ref:infn-napoli}}$,
L.\,Resmi$^{\ref{alma-inst:iisst}}$,
R.\,Ricci$^{\ref{alma-inst:bologna}}$
A.\,Rossi$^{\ref{rem-ref:oas}}$,
R.\,S\'anchez-Ram\'irez$^{\ref{iaps_roma}}$,
J.\,Selsing$^{\ref{bohr}}$
S.\,Schulze$^{\ref{alma-inst:weizmann}}$,
S.\,J.\,Smartt$^{\ref{rem-ref:QUB}}$, 
I.\,A.\,Smith$^{\ref{alma-inst:rice}}$,
V.\,V.\,Sokolov$^{\ref{inst:sao}}$,
J.\,Stevens$^{\ref{alma-inst:csiro}}$,
N.\,R.\,Tanvir$^{\ref{alma-inst:leicester}}$,
C.\,C.\,Th\'one$^{\ref{alma-inst:iaa}}$, 
A.\,Tiengo$^{\ref{xmm-inst:iuss},\ref{iasfmi}, \ref{infnpa}}$,
E.\,Tremou$^{\ref{inst:Sorbonne}}$,
E.\,Troja$^{\ref{inst:gsfc}, \ref{alma-inst:umd_astro}}$,
A.\,de\,Ugarte\,Postigo$^{\ref{alma-inst:iaa},\ref{alma-inst:dark}}$, A.\,F.\,Valeev$^{\ref{inst:sao}}$,
S.\,D.\,Vergani$^{\ref{inst:Meudon}}$,
M.\,Wieringa$^{\ref{alma-inst:atca}}$,
P.A.\,Woudt$^{\ref{inst:UCT}}$,
D. Xu$^{\ref{cas:china}}$,
O.\,Yaron$^{\ref{rem-ref:weizmann}}$, 
D.\,R.\,Young$^{\ref{rem-ref:QUB}}$ 
}

\begin{affiliations}  
\item \label{magic:1} Inst. de Astrof\'isica de Canarias, E-38200 La Laguna, and Universidad de La Laguna, Dpto. Astrof\'isica, E-38206 La Laguna, Tenerife, Spain 
\item \label{magic:2} Universit\`a di Udine, and INFN Trieste, I-33100 Udine, Italy 
\item \label{magic:31} Japanese MAGIC Consortium, Department of Physics, Kyoto University, Kyoto, Japan.
\item \label{magic:3} National Institute for Astrophysics (INAF), I-00136 Rome, Italy 
\item \label{magic:4} ETH Zurich, CH-8093 Zurich, Switzerland 
\item \label{magic:5} Technische Universit\"at Dortmund, 44221 Dortmund, Germany
\item \label{magic:36} Croatian Consortium, University of Zagreb, FER, Zagreb, Croatia.
\item \label{magic:7} Saha Institute of Nuclear Physics, HBNI, 1/AF Bidhannagar, Salt Lake, Sector-1, Kolkata 700064, India 
\item \label{magic:8} Centro Brasileiro de Pesquisas F\'isicas (CBPF), 22290-180 URCA, Rio de Janeiro (RJ), Brasil 
\item \label{magic:9} Unidad de Part\'iculas y Cosmolog\'ia (UPARCOS), Universidad Complutense, E-28040 Madrid, Spain 
\item \label{magic:10} University of Lodz, Faculty of Physics and Applied Informatics, Department of Astrophysics, 90-236 Lodz, Poland 
\item \label{magic:11} Universit\`a  di Siena and INFN Pisa, I-53100 Siena, Italy 
\item \label{magic:12} Deutsches Elektronen-Synchrotron (DESY), 15738 Zeuthen, Germany 
\item \label{magic:16} Universit\`a di Padova and INFN, I-35131 Padova, Italy 
\item \label{magic:13} Istituto Nazionale Fisica Nucleare (INFN), 00044 Frascati (Roma) Italy 
\item \label{magic:14} Max-Planck-Institut f\"ur Physik, 80805 M\"unchen, Germany 
\item \label{magic:15} Institut de F\'isica d'Altes Energies (IFAE), The Barcelona Institute of Science and Technology (BIST), E-08193 Bellaterra (Barcelona), Spain
\item \label{magic:17} Universit\`a di Pisa, and INFN Pisa, I-56126 Pisa, Italy 
\item \label{magic:40} The Armenian Consortium, A. Alikhanyan National Laboratory, Yerevan, Armenia.
\item \label{magic:ciemat} Centro de Investigaciones Energ\'{e}ticas, Medioambientales y Tecnol\'{o}gicas, E-28040 Madrid, Spain
\item \label{magic:26} also at Port d'Informaci\'o Cient\'ifica (PIC) E-08193 Bellaterra (Barcelona) Spain 
\item \label{magic:34} Croatian Consortium, Department of Physics, University of Rijeka, Rijeka, Croatia.
\item \label{magic:18} Universit\"at W\"urzburg, 97074 W\"urzburg, Germany 
\item \label{magic:41} Finnish MAGIC Consortium, Finnish Centre of Astronomy with ESO (FINCA), University of Turku, Turku, Finland.
\item \label{magic:20} Departament de F\'isica, and CERES-IEEC, Universitat Aut\`onoma de Barcelona, E-08193 Bellaterra, Spain 
\item \label{magic:30} Japanese MAGIC Consortium, ICRR, The University of Tokyo, Kashiwa, Japan. 
\item \label{magic:39} The Armenian Consortium, ICRANet-Armenia at NAS RA, Yerevan, Armenia.
\item \label{magic:35} Croatian Consortium, University of Split, FESB, Split, Croatia.
\item \label{magic:37} Croatian Consortium, Josip Juraj Strossmayer University of Osijek, Osijek, Croatia.
\item \label{magic:33} Japanese MAGIC Consortium, RIKEN, Wako, Japan.
\item \label{magic:32} Japanese MAGIC Consortium, Tokai University, Hiratsuka, Japan
\item \label{magic:28} Istituto Nazionale di Fisica Nucleare, Sezione di Trieste, 34149 Trieste, Italy 
\item \label{magic:25} also at Dipartimento di Fisica, Universit\`a di Trieste, 34127 Trieste, Italy 
\item \label{magic:23} Inst. for Nucl. Research and Nucl. Energy, Bulgarian Academy of Sciences, BG-1784 Sofia, Bulgaria 
\item \label{magic:24} Universitat de Barcelona, ICCUB, IEEC-UB, E-08028 Barcelona, Spain  
\item \label{magic:42} Finnish MAGIC Consortium, Astronomy Research Unit, University of Oulu, Oulu, Finland.
\item \label{magic:38} Croatian Consortium, Rudjer Boskovic Institute, Zagreb, Croatia.
\item \label{magic:29} Institute for Fundamental Physics of the Universe (IFPU), 34151 Trieste, Italy 
\item \label{fermi-inst:hsv} Center for Space Plasma and Aeronomic Research, University of Alabama in Huntsville, 320 Sparkman Drive, Huntsville, AL 35899, USA 
\item \label{fermi-inst:hsv2} Space Science Department, University of Alabama in Huntsville, 320 Sparkman Drive, Huntsville, AL 35899, USA 
\item \label{fermi-inst:usra} Science and Technology Institute, Universities Space Research Association, Huntsville, AL 35805, USA
\item \label{fermi-inst:msfc} Astrophysics Office, ST12, NASA/Marshall Space Flight Center, Huntsville, AL 35812, USA 
\item \label{fermi-inst:mun} Max-Planck Institut f\"ur extraterrestrische Physik, Giessenbachstra{\ss}e 1, 85748 Garching, Germany 
\item \label{inst:Kanazawa} Faculty of Mathematics and Physics, Institute of Science and Engineering, Kanazawa University, Kakuma, Kanazawa, Ishikawa 920-1192 
\item \label{alma-inst:umd_phys} Department of Physics, University of Maryland, College Park, MD 20742-4111, USA 
\item \label{inst:gsfc} Astrophysics Science Division, NASA Goddard Space Flight Center, 8800 Greenbelt Rd, Greenbelt, MD 20771, USA 
\item \label{fermi-inst:ICRR} Institute for Cosmic-Ray Research, University of Tokyo, 5-1-5 Kashiwanoha, Kashiwa, Chiba, 277-8582, Japan 
\item \label{fermi-inst:SU} Department of Physics, Stockholm University, AlbaNova, SE-106 91 Stockholm, Sweden 
\item \label{fermi-inst:KTH} Department of Physics, KTH Royal Institute of Technology, AlbaNova, SE-106 91 Stockholm, Sweden 
\item \label{fermi-inst:UbiBa} Dipartimento di Fisica ``M. Merlin" dell'Universit\`a e del Politecnico di Bari, 70126 Bari, Italy 
\item \label{fermi-inst:INFNBa} Istituto Nazionale di Fisica Nucleare, Sezione di Bari, 70126 Bari, Italy 
\item \label{fermi-inst:UniJoh} Department of Physics, University of Johannesburg, PO Box 524, Auckland Park 2006, South Africa 
\item \label{fermi-inst:UniIsrael} Department of Natural Sciences, Open University of Israel, 1 University Road, POB 808, Ra'anana 43537, Israel 
\item \label{inst:stanford1} W.\ W.\ Hansen Experimental Physics Laboratory, Kavli Institute for Particle Astrophysics and Cosmology.
\item \label{inst:stanford2} Department of Physics and SLAC National Accelerator Laboratory, Stanford University, Stanford, CA 94305, USA  
\item \label{fermi-inst:LUPM} Laboratoire Univers et Particules de Montpellier, Universit\'e Montpellier, CNRS/IN2P3, F-34095 Montpellier, France
\item \label{rem-ref:oab}  INAF - Astronomical Observatory of Brera, I-23807 Merate (LC), Italy  
\item \label{uvot-inst:mssl} Mullard Space Science Laboratory, University College London, Holmbury St. Mary, Dorking, RH5 6NT, United Kingdom 
\item \label{uvot-inst:psu} Department of Astronomy and Astrophysics, Pennsylvania State University. 525 Davey Laboratory, University Park, PA 16802, USA   
\item \label{cenko2} Joint Space-Science Institute, University of Maryland, College Park, Maryland 20742, USA 
\item \label{alma-inst:leicester} Department of Physics and Astronomy, University of Leicester, University Road, Leicester LE1 7RH, UK 
\item \label{palermo} INAF Istituto di Astrofisica Spaziale e Fisica Cosmica di Palermo, via Ugo La Malfa 153, I-90146 Palermo, Italia 
\item \label{turkey} Department of Astronomy and Space Sciences, Istanbul University, Fatih, 34119, Istanbul, Turkey
\item \label{OAR} INAF-Osservatorio Astronomico di Roma, Via Frascati 33, I-00078 Monteporzio Catone, Italy 
\item \label{SSDC} Space Science Data Center (SSDC), Agenzia Spaziale Italiana (ASI), via del Politecnico s.n.c., I-00133, Roma, Italy 
\item \label{iaps_roma} INAF-IAPS, via del Fosso del Cavaliere 100, I-00133 Roma, Italy 
\item \label{agile2} Univ. ``Tor Vergata'', Via della Ricerca Scientifica 1, I-00133 Roma, Italy 
\item \label{agile3} Gran Sasso Science Institute, viale Francesco Crispi 7, I-67100 L'Aquila, Italy 
\item \label{alma-inst:icra} International Centre for Radio Astronomy Research, Curtin University, GPO Box U1987, Perth, WA 6845, Australia
\item \label{alma-inst:eso} European Southern Observatory, Alonso de C\`ordova, 3107, Vitacura, Santiago 763-0355, Chile 
\item \label{alma-inst:bologna} INAF-Istituto di Radioastronomia, via Gobetti 101, I-40129, Bologna, Italy 
\item \label{alma-inst:rsa} Department of Physics and Electronics, Rhodes University, PO Box 94, Grahamstown, 6140, South Africa 
\item  \label{alma-inst:ska} South African Radio Astronomy Observatory, Black River Park, 2 Fir Street, Observatory, Cape Town, 7925, South Africa 
\item \label{inst:asu} Astronomical Institute of the Academy of Sciences, Bo\v cn\'i II 1401, CZ-14100 Praha 4, Czech Republic 
\item \label{inst:uma3} Departamento de F\'isica Aplicada, Facultad de Ciencias, Universidad de M\'alaga, Bulevar Louis Pasteur 31, E-29071 M\'alaga, Spain 
\item \label{inst:uma2} Departamento de \'Algebra, Geometr\'ia y Topolog\'ia, Facultad de Ciencias, Bulevar Louis Pasteur 31, Universidad de M\'alaga, E-29071 M\'alaga, Spain 
\item \label{rem-ref:soton} Physics and Astronomy Department, University of Southampton, Southampton, UK
\item \label{inst:uma} Unidad Asociada Departamento de Ingenier\'ia de Sistemas y Autom\'atica, E.T.S. de Ingenieros Industriales, Universidad de M\'alaga, Arquitecto Francisco Pe\~nalosa 6, E-29071 M\'alaga, Spain 
\item \label{alma-inst:iaa} Instituto de Astrof\' isica de Andaluc\' ia (IAA-CSIC), Glorieta de la Astronom\' ia, s/n, E-18008, Granada, Spain 
\item \label{alma-inst:iisst} Indian Institute of Space Science \& Technology, Trivandrum 695547, India 
\item \label{alma-inst:LivJMU} Astrophysics Research Institute, Liverpool John Moores University, 146 Brownlow Hill, Liverpool L3 5RF, UK 
\item \label{rem-ref:oasdg} Osservatorio Astronomico 'S. Di Giacomo' - AstroCampania, I-80051, Agerola (NA), Italy 
\item \label{rem-ref:oac} INAF - Astronomical Observatory of Naples, I-23807 Naples (NA), Italy
\item \label{inst:UCT} Inter-University Institute for Data-Intensive Astronomy, Department of Astronomy, University of Cape Town, Private Bag X3, Rondebosch 7701, South Africa 
\item \label{inst:Oxford} Astrophysics, Department of Physics, University of Oxford, Keble Road, Oxford OX1 3RH, UK
\item \label{dawn}  Cosmic Dawn Center (DAWN)
\item \label{bohr} Niels Bohr Institute, Copenhagen University, Vibenshuset, Lyngbyvej 2, DK-2100, Copenhagen
\item \label{baltimore} Space Telescope Science Institute, 3700 San Martin Drive, Baltimore, MD 21218, USA 
\item \label{rem-ref:warsaw} Astronomical Observatory, University of Warsaw, Al. Ujazdowskie 4, PL-00- 478 Warszawa, Poland
\item \label{iceland} Centre for Astrophysics and Cosmology, Science Institute, University of Iceland, Dunhagi 5, 107 Reykjav\'ik, Iceland 
\item \label{inst:GWU} Department of Physics, The George Washington University, 725 21st Street NW, Washington, DC 20052, USA 
\item \label{inst:APSIS} Astronomy, Physics, and Statistics Institute of Sciences (APSIS), The George Washington University, Washington, DC 20052, USA 
\item \label{inst:ugr} Universidad de Granada, Facultad de Ciencias Campus Fuentenueva S/N CP 18071 Granada, Spain 
\item \label{rem-ref:cardiff} School of Physics \& Astronomy, Cardiff University, Queens Buildings, The Parade, 25 Cardiff, CF24 3AA, UK
\item \label{alma-inst:dark} DARK, Niels Bohr Institute, University of Copenhagen, Lyngbyvej 2, DK-2100 Copenhagen \O, Denmark 
\item \label{pannekoek} Anton Pannekoek Institute for Astronomy, University of Amsterdam, Science Park 904, 1098 XH Amsterdam, The Netherlands
\item \label{rem-ref:turku} Tuorla Observatory, Department of Physics and Astronomy, University of Turku, 20014 Turku, Finland
\item \label{alma-inst:tls} Th\"uringer Landessternwarte Tautenburg, Sternwarte 5, 07778 Tautenburg, Germany 
\item \label{nijmegen} Department of Astrophysics/IMAPP, Radboud University Nijmegen, The Netherlands 
\item \label{inst:ihsm} Instituto de Hortofruticultura Subtropical y Mediterr\'anea La Mayora (IHSM/UMA-CSIC), Algarrobo Costa s/n, E-29750 M\'alaga, Spain 
\item \label{inst:niaot} Nanjing Institute for Astronomical Optics and Technology, National Observatories, Chinese Academy of Sciences, 188 Bancang St, Xuanwu Qu, Nanjing Shi, Jiangsu Sheng, China 
\item \label{rem-ref:trinity} School of Physics, Trinity College Dublin, Dublin 2, Ireland
\item \label{malesani1} DTU Space, National Space Institute, Technical University of Denmark, Elektrovej 327, 2800 Kongens Lyngby, Denmark
\item \label{rem-ref:weizmann} Benoziyo Center for Astrophysics, Weizmann Institute of Science, 76100 Rehovot, Israel
\item \label{alma-inst:ferrara} Department of Physics and Earth Science, University of Ferrara, via Saragat 1, I-44122, Ferrara, Italy \item \label{icranet}  International Center for Relativistic Astrophysics Network (ICRANet), Piazzale della Repubblica 10, I-65122, Pescara, Italy 
\item \label{alma-inst:alma} Joint ALMA Observatory, Alonso de C\`ordova, 3107, Vitacura, Santiago 763-0355, Chile 
\item \label{alma-inst:poz} Astronomical Observatory Institute, Faculty of Physics, Adam Mickiewicz University, ul.~S{\l}oneczna 36, 60-286 Pozna{\'n}, Poland 
\item \label{alma-inst:aries} Aryabhatta Research Institute of Observational Sciences, Manora Peak, Nainital 263 001, India  
\item \label{misra2} Department of Physics, University of California, 1 Shields Ave, Davis, CA 95616-5270, USA 
\item \label{alma-inst:uaeu} Physics Department, United Arab Emirates University, P.O. Box 15551, Al-Ain, United Arab Emirates 
\item \label{inst:NRAO} National Radio Astronomy Observatory, 1003 Lopezville Road, Socorro, NM 87801, USA 
\item \label{inst:Caltech} Caltech, 1200 California Blvd., Pasadena, CA 91106, USA 
\item \label{rem-ref:edinburgh} Institute for Astronomy, University of Edinburgh, Royal Observatory, Blackford Hill, EH9 3HJ, UK
\item \label{rem-ref:birmingham} Birmingham Institute for Gravitational Wave Astronomy and School of Physics and Astronomy, University of Birmingham, Birmingham B15 2TT, UK
\item \label{xmm-inst:iuss} Scuola Universitaria Superiore IUSS Pavia, Piazza della Vittoria 15, 27100 Pavia, Italy 
\item \label{iasfmi} INAF - IASF Milano, Via E. Bassini 15, 20133 Milano, Italy 
\item \label{infn-lngs} INFN / Laboratori Nazionali del Gran Sasso, Via G. Acitelli 22, 67100, Assergi (AQ), Italy 
\item \label{inst:Zaragoza} Depto. de F\'isica Te\'orica, Universidad de Zaragoza, E-50019, Zaragoza, Spain 
\item \label{rem-ref:uni-napoli} Dipartimento di Scienze Fisiche, Università degli studi di Napoli Federico II, Via Cinthia, Edificio N, 80126 Napoli, Italy
\item \label{rem-ref:infn-napoli} INFN, Sezione di Napoli, Complesso Universitario di Monte S. Angelo, Via Cinthia, Edificio N, 80126 Napoli, Italy
\item \label{rem-ref:oas} INAF - Osservatorio di Astrofisica e Scienza dello Spazio, via Piero Gobetti 93/3, 40129 Bologna, Italy
\item \label{alma-inst:weizmann} Department of Particle Physics and Astrophysics, Weizmann Institute of Science, Rehovot 76100, Israel 
\item \label{rem-ref:QUB} Astrophysics Research Centre, School of Mathematics and Physics, Queen?s University Belfast, Belfast BT7 1NN, UK
\item \label{alma-inst:rice} Department of Physics and Astronomy, Rice University, 6100 South Main, MS-108, Houston, TX 77251-1892, USA 
\item \label{inst:sao} Special Astrophysical Observatory, Nizhniy Arkhyz, Zelenchukskiy region, Karachai-Cherkessian Republic, 369167, Russia 
\item \label{alma-inst:csiro} CSIRO Australia Telescope National Facility, Paul Wild Observatory, Narrabri NSW 2390, Australia 
\item \label{infnpa} Istituto Nazionale di Fisica Nucleare, Sezione di Pavia, Via Bassi 6,
27100 Pavia, Italy 
\item \label{inst:Sorbonne} AIM, CEA, CNRS, Universit\'{e} Paris Diderot, Sorbonne Paris Cit\'{e}, Universit\'{e} Paris-Saclay, F-91191 Gif-sur-Yvette, France 
\item \label{alma-inst:umd_astro} Department of Astronomy, University of Maryland, College Park, MD 20742-4111, USA 
\item \label{inst:Meudon} GEPI, Observatoire de Paris, PSL University, CNRS, 5 Place Jules Janssen, 92190 Meudon, France 
\item \label{alma-inst:atca} Australia Telescope National Facility, CSIRO Astronomy and Space Science, PO Box 76, Epping, NSW 1710, Australia 
\item \label{cas:china} CAS Key Laboratory of Space Astronomy and Technology, National Astronomical Observatories, Chinese Academy of Sciences, Beijing 100012, China 
\end{affiliations}
\end{centering}

\begin{abstract}
Long-duration gamma-ray bursts (GRBs) originate from ultra-relativistic jets launched from the collapsing cores of dying massive stars. They are characterised by an initial phase of bright and highly variable radiation in the keV-MeV band that is likely produced within the jet and lasts from milliseconds to minutes, known as the prompt emission\cite{Meszaros2002,Piran2004}. Subsequently, the interaction of the jet with the external medium  generates external shock waves, responsible for the afterglow emission, which lasts from days to months, and occurs over a broad energy range, from the radio to the GeV bands\cite{vanParadijsetal2000,Meszaros2002,Piran2004,Gehrelsetal2009,Gehrels&Meszaros2012,Kumar&Zhang2015}. 
The afterglow emission is generally well explained as synchrotron radiation by electrons accelerated at the external shock\cite{Sarietal1998,Granot&Sari2002,Meszaros&Rees1994}.
Recently, an intense, long-lasting emission between 0.2 and 1\,TeV was
observed from the GRB 190114C\cite{DiscoveryPaper}. Here we present the results of our multi-frequency observational campaign of GRB~190114C, and study the evolution in time of the GRB emission across 17 orders of magnitude in energy, from $5\times10^{-6}$ up to $10^{12}$\,eV. 
We find that the broadband spectral energy distribution is double-peaked, with the TeV emission constituting a distinct spectral component that has power comparable to the synchrotron component. This component is associated with the afterglow, and is satisfactorily explained by inverse Compton upscattering of synchrotron photons by high-energy electrons. 

We find that the conditions required to account for the observed TeV component are not atypical, supporting the possibility that inverse Compton emission is commonly produced in GRBs.
\end{abstract}

On 14 January 2019, following an alert from the Neil Gehrels Swift Observatory (hereafter {\it Swift}) and the {\it Fermi} satellite, the Major Atmospheric Gamma Imaging Cherenkov (MAGIC) telescopes observed and detected radiation up to at least 1\,TeV from GRB\,190114C. 
Before the MAGIC detection, GRB emission has only been reported at much lower energies, $\lesssim100$\,GeV, first by {\it CGRO}/EGRET in a handful of cases, and more recently by {\it AGILE}/GRID and {\it Fermi}/LAT (see \cite{Nava2018} for a recent review). 

\begin{figure}
\centering
\includegraphics[width=0.93\textwidth]{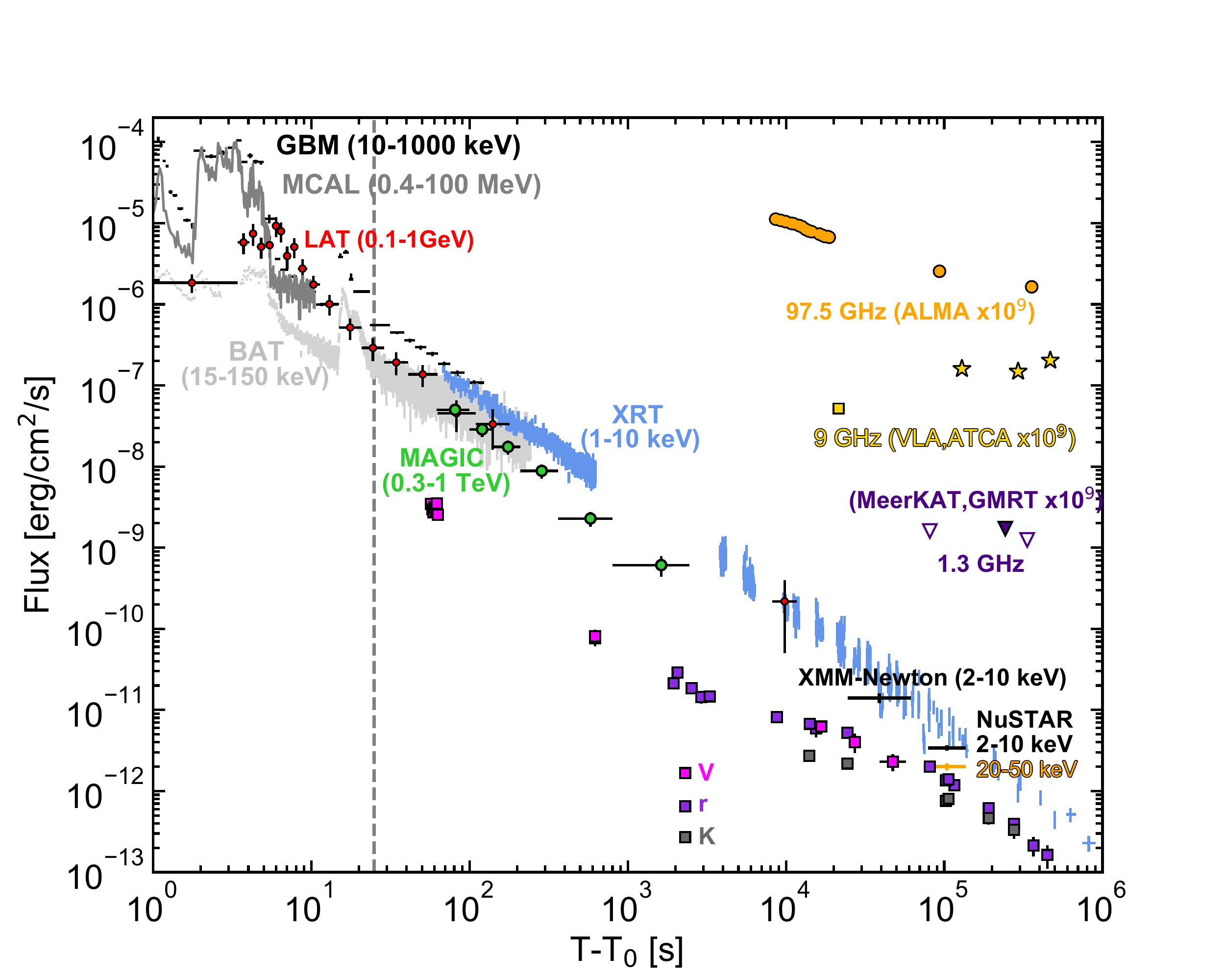}
\caption{{\bf Multi-wavelength light curves of GRB\,190114C.}
Energy flux at different wavelengths, from radio to gamma-rays, versus time since the BAT trigger time $T_0$\,=\,20:57:03.19\,UT on 14 January 2019.  The light curve for the energy range 0.3-1\,TeV
(green circles) is compared with light curves at lower frequencies.
Those for VLA (yellow square), ATCA (yellow stars), ALMA (orange circles), GMRT (purple filled triangle), and MeerKAT (purple empty triangles) have been multiplied by $10^9$ for clarity.
The vertical dashed line marks approximately the end of the prompt emission phase, identified with the end of the last flaring episode.
For the data points, vertical bars show the 1-$\sigma$ errors on the flux, while horizontal bars represent the duration of the observation.
}
\label{fig:lc}
\end{figure}

Detection of TeV radiation opens a new window in the electromagnetic spectrum for the study of GRBs\cite{DiscoveryPaper}.
Its announcement\cite{Mirzoyan2019} triggered an extensive campaign of follow-up observations.
Owing to the relatively low redshift $z=0.4245\pm0.0005$ (see Methods) of the GRB (corresponding to a luminosity distance of $\sim2.3$\,Gpc)
a comprehensive set of multi-wavelength data could be collected.
We present observations gathered from instruments onboard six satellites and 15 ground telescopes (radio, submm and NIR/optical/UV and very high energy gamma-rays; see Methods) for the first ten days after the burst.
The frequency range covered by these observations spans more than 17 orders of magnitude, from 1 to $\sim2\times10^{17}$\,GHz, the most extensive to date for a GRB.
The light curves of GRB~190114C at different frequencies are shown in Fig.~\ref{fig:lc}.

The prompt emission of GRB~190114C was simultaneously observed by several space missions (see Methods), covering the spectral range from 8\,keV to $\sim$\,100\,GeV.
The prompt light curve shows a complex temporal structure, with several emission peaks (Methods; Extended Data Fig.~\ref{fig:prompt_lc}), with total duration $\sim$\,25\,s (see dashed line in Fig.~\ref{fig:lc}) and total radiated energy $E_{\rm \gamma,iso}$ = (2.5 $\pm$ 0.1) $\times10^{53}$ ergs  (isotropic equivalent, in the energy range $1-10^4$\,keV \cite{FermiSwift19}).
During the time of inter-burst quiescence at $t \sim [5-15]$ seconds and after the end of the last prompt pulse at $t\gtrsim 25$\,s, the flux decays smoothly, following a power law in time $F\propto t^\alpha$, with $\alpha_{10-1000{\rm keV}}=-1.10\pm 0.01$ \cite{FermiSwift19}.
The temporal and spectral characteristics of this smoothly varying component support an interpretation in terms of afterglow synchrotron radiation, making this one of the few clear cases of afterglow emission detected in the band $10-10^4$\,keV during the prompt emission phase.
The onset of the afterglow component is then estimated to occur around $ t \sim 5-10$\,s \cite{ravasioetal2019,FermiSwift19}, implying an initial bulk Lorentz factor between 300 and 700 (Methods). 

After about one minute from the start of the prompt emission, two additional high-energy telescopes began observations:
MAGIC and {\it Swift}/XRT.
The XRT and MAGIC light curves (1-10\,keV, blue data points in Fig.~\ref{fig:lc}, and 0.3-1\,TeV, green data points, respectively) decay with time as a power law, and display the following decay rates: $\alpha_{\rm X}\sim -1.36\pm0.02$ and $\alpha_{\rm TeV}\sim-1.51\pm0.04$. 
The 0.3-1\,TeV light curve shown in Fig.~\ref{fig:lc} was obtained after correcting for attenuation by the extragalactic background light (EBL)\cite{DiscoveryPaper}.
The TeV-band emission is observable until $\sim$\,40 minutes, which is much longer than the nominal duration of the prompt emission phase.
The NIR-optical light curves (square symbols) show a more complex behaviour. Initially, a fast decay is seen, where the emission is most likely dominated by the reverse shock component\cite{Laskaretal2019}. This is followed by a  shallower decay, and subsequently a faster decay at $t\gtrsim10^5$\,s. The latter behaviour might indicate that the characteristic synchrotron frequency \nui\ is crossing the optical band (Extended Data Fig.~\ref{fig:SED_late}), which is not atypical, but usually occurs at earlier times.
The relatively late time at which the break appears in GRB~190114C would then imply a very large value of $\nu_{\rm m}$, placing it in the X-ray band at $\sim10^2$\,s.
The millimeter light curves (orange symbols) also show an initial fast decay where the emission is dominated by the reverse shock, followed by emission at late times with nearly constant flux (Extended Data Fig.~\ref{fig:radio_lc}).

\begin{figure}
\centering
\includegraphics[width=0.9\textwidth]{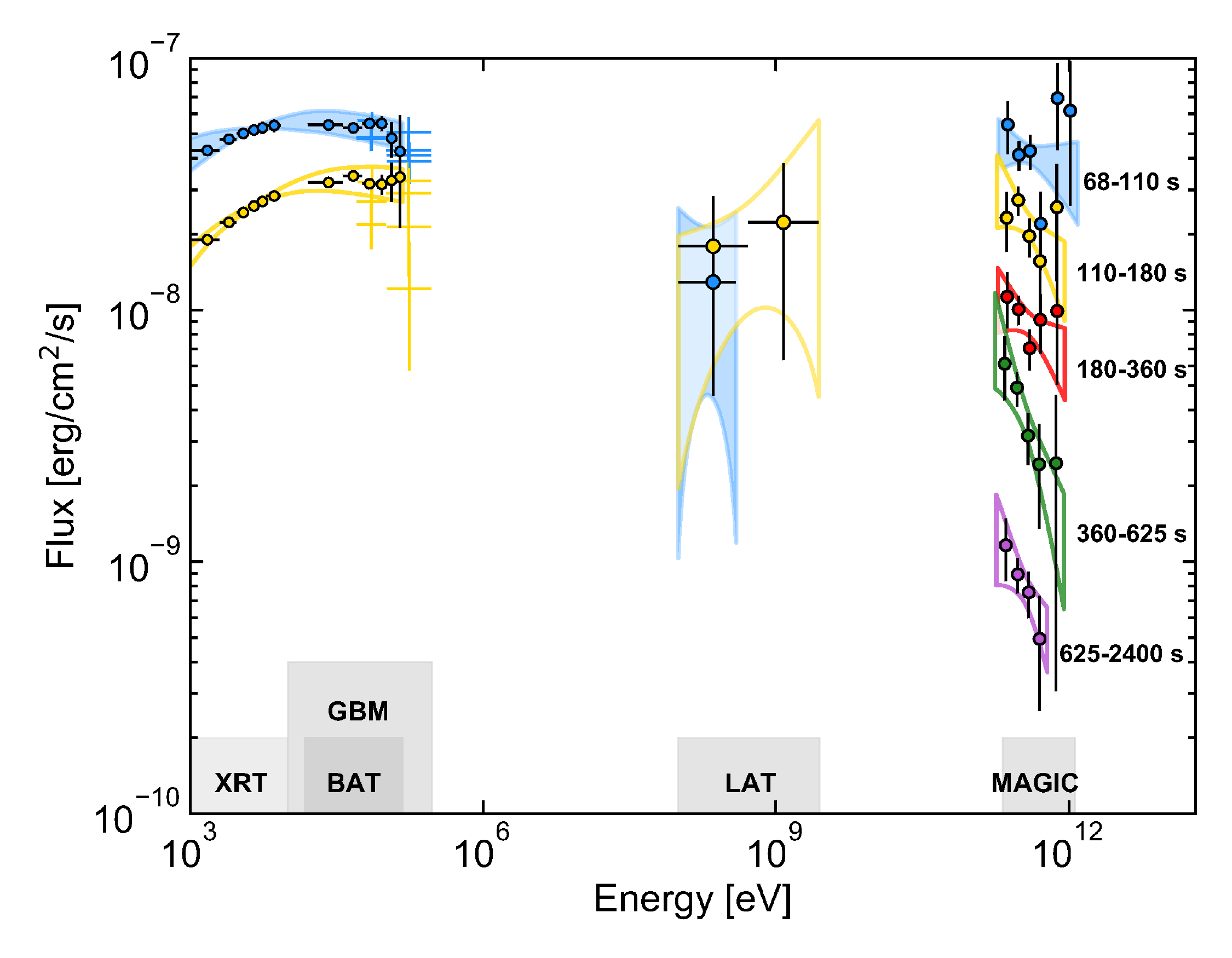}
\caption{{\bf Multi-band spectra in the time interval 68-2400\,s}. Five time intervals are considered: 68-110\,s (blue), 110-180\,s (yellow), 180-360\,s (red), 360-625\,s (green), 625-2400\,s (purple).
MAGIC data points have been corrected for attenuation caused by the extragalactic background light.
Data from other instruments are shown for the first two time-intervals: {\it Swift}/XRT, {\it Swift}/BAT, {\it Fermi}/GBM, and {\it Fermi}/LAT. For each time interval, LAT contour regions are shown limiting the energy range to the range where photons are detected.
MAGIC and LAT contour regions are drawn from the 1-$\sigma$ error of their best-fit power law functions.
For {\it Swift} data, the regions show the 90\% confidence contours for the joint fit XRT-BAT obtained fitting to the data a smoothly broken power law.
Filled regions are used for the first time interval (68-110\,s, blue color).
}
\label{fig:SED_MAGIC_all}
\end{figure}

The spectral energy distributions (SEDs) of the radiation detected by MAGIC are shown in Fig.~\ref{fig:SED_MAGIC_all}, where the whole duration of the emission detected by MAGIC is divided into five time intervals.
For the first two time intervals, observations in the GeV and X-ray bands are also available.
During the first time interval (68-110\,s, blue data points and blue confidence regions), {\it Swift}/XRT-BAT and {\it Fermi}/GBM data show that the afterglow synchrotron component is peaking in the X-ray band.
At higher energies, up to $\lesssim$1\,GeV, the SED is a decreasing function of energy, as supported by the {\it Fermi}/LAT flux between 0.1 and 0.4\,GeV (see Methods).
On the other hand, at even higher energies, the MAGIC flux above 0.2\,TeV implies a spectral hardening. This evidence is independent of the EBL model adopted to correct for the attenuation (Methods).
This demonstrates that the newly discovered TeV radiation is not a simple extension of the known afterglow synchrotron emission, but rather a separate spectral component that has 
never been clearly seen before.

The extended duration and the smooth, power-law temporal decay of the radiation detected by MAGIC (see green data points in Fig.~\ref{fig:lc}) suggest an intimate connection between the TeV emission and the broadband afterglow emission.
The most natural candidate is synchrotron self-Compton (SSC) radiation in the external forward shock:
the same population of relativistic electrons responsible for the afterglow synchrotron emission Compton upscatters the synchrotron photons, leading to a second spectral component that peaks at higher energies.
TeV afterglow emission can also be produced by hadronic processes such as synchrotron radiation by protons accelerated to ultra-high energies in the forward shock\cite{Vietri1997,Zhang&Meszaros2001,Razzaque2010}.
However, due to their typically low efficiency of radiation\cite{Kumar&Zhang2015}, reproducing the luminous TeV emission as observed here by such processes would imply unrealistically large power in accelerated protons\cite{DiscoveryPaper}.
TeV photons can also be produced via the SSC mechanism in internal shock synchrotron models of the prompt emission.
However, numerical modeling (Methods) shows that prompt SSC radiation can account at most for a limited fraction ($\lae$20$\%$) of the observed TeV flux, and only at early times ($t \lesssim100$\,s).
Henceforth, we focus on the SSC process in the afterglow.

SSC emission has been 
predicted for GRB afterglows\cite{Meszaros&Rees1994,Sari&Esin2001,Zhang&Meszaros2001,Meszarosetal2004,Lemoine2015,Fan&Piran2008,Galli&Piro2008,Nakaretal2009,Xueetal2009,Piran&Nakar2010,Nava2018}. 
However, 
its
quantitative significance was uncertain, as 
the SSC
luminosity and spectral properties depend strongly on the poorly constrained physical conditions in the emission region (e.g., the magnetic field strength).
The
detection of the TeV component in GRB~190114C and the availability of multi-band observations offer the 
opportunity to investigate the relevant physics at a deeper level. 
SSC radiation might have been already detected in very bright GRBs, such as GRB~130427A. Photons with energies 10-100\,GeV, as those detected in GRB~130427A are challenging to explain by the synchrotron processes, suggesting a different origin\cite{Tametal2013,Liuetal2013,FermiGRB130427A}.

\begin{figure}
\centering
\includegraphics[width=0.9\textwidth]{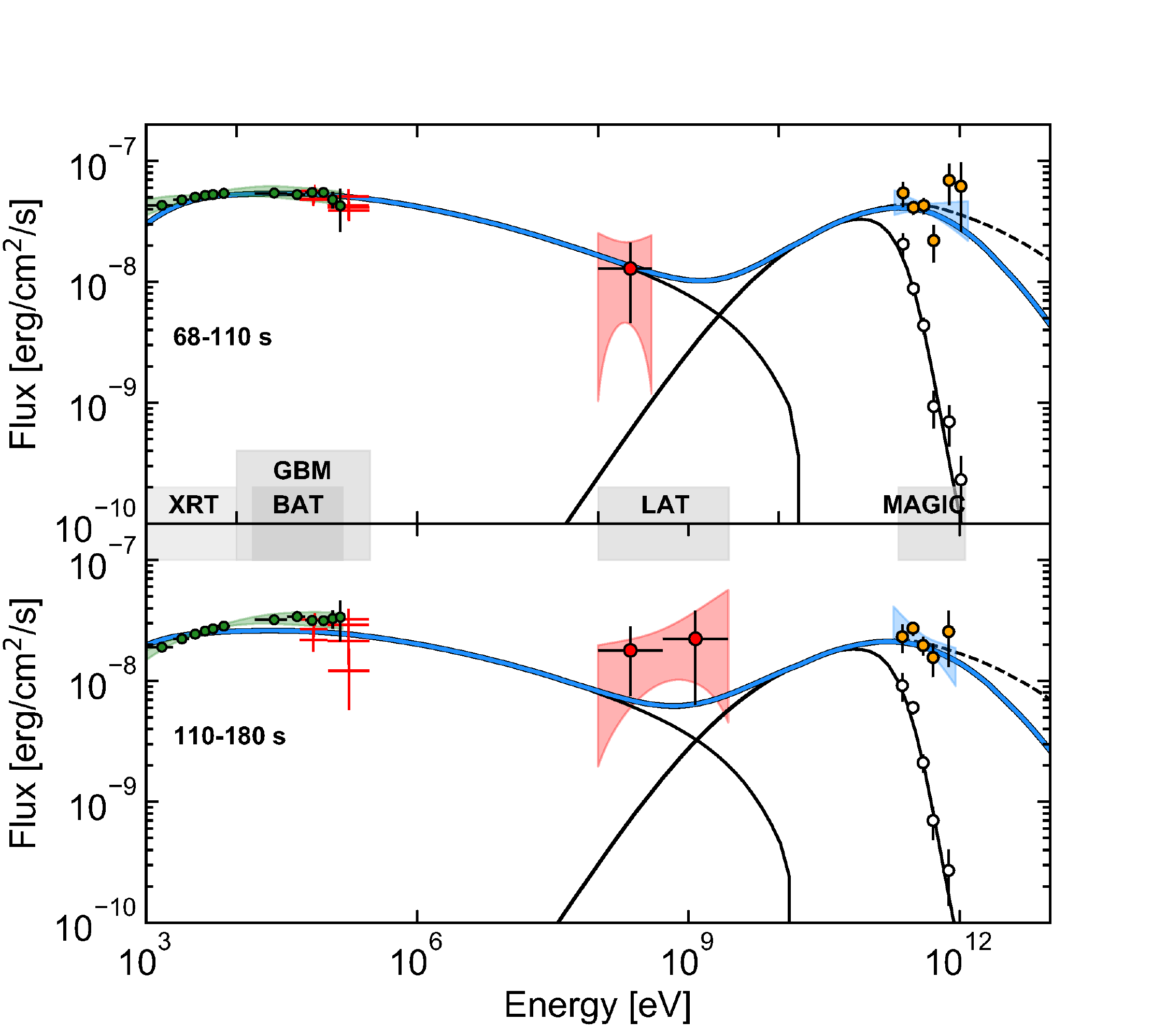}
\caption{{\bf
Modeling of the broadband spectra in the time intervals 68-110\,s and 110-180\,s.}
Thick blue curve: modeling of the multi-band data in the synchrotron and SSC afterglow scenario. 
Thin solid lines: synchrotron and SSC (observed spectrum) components; dashed lines: SSC if internal \pp\ opacity is neglected.
The adopted parameters are: $s=0$, $\epsilon_{\rm e}=0.07$, $\epsilon_{\rm B}=8\times10^{-5}$, $p=2.6$, $n_0=0.5$, and $E_{\rm k}=8\times10^{53}$\,erg, see the Text.
Empty circles show the observed MAGIC spectrum, i.e. not corrected by attenuation caused by the extragalactic background light. Contour regions and data points as in Fig.~\ref{fig:SED_MAGIC_all}.
}
\label{fig:SED_68-180}
\end{figure}

We model the full data set (from radio band to TeV energies, for the first week after the explosion) as synchrotron plus SSC radiation, within the framework of the theory of afterglow emission from external forward shocks. The detailed modeling of the broadband emission and its evolution with time is presented in Section Methods. We discuss here the implications for the  emission at $t<2400$\,s and energies above $>1$\,keV.

The soft spectra in the 0.2-1\,TeV energy range (photon index $\Gamma_{\rm TeV}<-2$; see Extended Data Table~\ref{tab:magicSfit}) constrain the peak of the SSC component to be below this energy range. 
The relatively small ratio between the spectral peak energies of the SSC ($E_{\rm p}^{\rm SSC}\lesssim 200$\,GeV) and synchrotron ($E_{\rm p}^{\rm syn}\sim 10$\,keV) components implies a relatively low value for the Lorentz factor of the electrons ($\gamma\sim2\times10^3$).
This value is hard to reconcile with the observation of the synchrotron peak at $\gtrsim$\,keV energies. 
In order to explain the soft spectrum detected by MAGIC, it is necessary to invoke the Klein-Nishina (KN) regime scattering for the electrons radiating at the spectral peak as well as internal \pp\ absorption\cite{WangZhang2019}.
While both effects tend to become less important with time, the spectral index in the 0.2-1\,TeV band remains constant in time (or possibly evolves to softer values; Extended Data Table~\ref{tab:magicSfit}). This implies that the SSC peak energy is moving to lower energies  and crossing the MAGIC energy band.
The energy at which attenuation by internal pair production becomes important indicates that the bulk Lorentz factor is $\sim$140-160 at $100$\,s.

An example of the theoretical modeling  in this scenario is shown in  Fig.~\ref{fig:SED_68-180} (blue solid curve, see Methods for details).
The dashed line shows the SSC spectrum when internal absorption is neglected. The thin solid line shows the  model spectrum including EBL attenuation, in comparison to the MAGIC observations (empty circles).

We find that acceptable models of the broadband SED can be obtained if the conditions at the source are the following. The initial kinetic energy of the blastwave is $E_{\rm k}\gtrsim3\times10^{53}$\,erg (isotropic-equivalent). 
The electrons swept up from the external medium are efficiently injected into the acceleration process, and carry a fraction \ee\,$\sim0.05-0.15$ of the energy dissipated at the shock. The acceleration mechanism produces an electron population characterized by a non-thermal energy distribution, described by a power law with index $p\sim2.4-2.6$, injection Lorentz factor $\gamma_{\rm m}=(0.8-2)\times10^4$ and maximum Lorentz factor $\gamma_{\max} \sim10^8$ (at $\sim100$\,s).
The magnetic field behind the shock conveys a fraction \eb\,$\sim(0.05-1) \times10^{-3}$ of the dissipated energy.
At $t\sim100$\,s, corresponding to $R\sim (8-20)\times10^{16}$\,cm, the density of the external medium is $n\sim0.5-5$\,cm$^{-3}$, and the magnetic field strength is $B \sim 0.5-5$\,Gauss.
The latter implies that the magnetic field was efficiently amplified from values of a few $\mu$Gauss that are typical of the unshocked ambient medium, due to plasma instabilities or other mechanisms\cite{Kumar&Zhang2015}.
 Not surprisingly, we find that $\epsilon_{\rm e}\gg\epsilon_{\rm B}$, that is a necessary condition for the efficient production of SSC radiation\cite{Sari&Esin2001,Zhang&Meszaros2001}.

The blastwave energy inferred from the modeling is comparable to
the amount of energy released in the form of radiation during the prompt phase. 
The prompt emission mechanism must then have dissipated and radiated no more than half of the initial jet energy, leaving the other half available for the afterglow phase.
The modeling of the multi-band data also allows us to infer how the total energy is shared between the synchrotron and the SSC components.
The resultant power in the two components are comparable. 
We estimate that the energy in the synchrotron and SSC component are $\sim1.5\times10^{52}$\,erg and $\sim6.0\times10^{51}$\,erg respectively in the time interval 68-110\,s, and $\sim1.3\times10^{52}$\,erg and $\sim5.4\times10^{51}$\,erg respectively in the time interval 110-180\,s.
Thus, previous studies of GRBs may have been missing a significant fraction of the energy emitted during the afterglow phase that is essential to its understanding.

Finally, we note that the values of the afterglow parameters inferred from the modeling fall within the range of values typically inferred from broadband (radio-to-GeV) studies of GRB afterglow emission.
This points to the possibility that SSC emission in GRBs may be a relatively common process that does not require special conditions to be produced with power similar to synchrotron radiation.

The SSC component may then be detectable at TeV energies in other relatively energetic GRBs, as long as the redshift is low enough to avoid severe attenuation by the EBL. This also provides support to earlier indications for SSC emission at GeV energies \cite{Tametal2013,Liuetal2013,FermiGRB130427A}.

\newpage
\begin{methods}\label{sec:methods}

\renewcommand{\figurename}{Extended Data Figure}
\setcounter{figure}{0}
\renewcommand{\tablename}{Extended Data Table}

\subsection{Prompt emission observations}
On 14 January 2019, the prompt emission from GRB\,190114C triggered several space instruments, including
{\it Fermi}/GBM\cite{Hamburg2019}, {\it Fermi}/LAT\cite{LATGCN}, {\it Swift}/BAT\cite{Gropp2019},  
Super-AGILE\cite{GCNAgile}, 
 {\it AGILE}/MCAL\cite{GCNAgile}, 
 {\it KONUS}/Wind\cite{GCNKonus}, 
 {\it INTEGRAL}/SPI-ACS\cite{GCNIntegral},  and  
{\it Insight}/HXMT\cite{GCNInsight}.
The prompt emission light curves from {\it AGILE}, {\it Fermi}, and {\it Swift} are shown in Fig.~\ref{fig:lc} and in Extended Data Fig.~\ref{fig:prompt_lc},
where the trigger time $T_0$ (here and elsewhere) refers to the BAT trigger time (20:57:03.19 UT). 
The prompt emission lasts approximately for 25\,s, where the last flaring emission episode ends.
Nominally, the $T_{90}$, i.e. the time interval during which a fraction between 5\% and 95\% of the total emission is observed,
is much longer ($>100$\,s, depending on the instrument \cite{FermiSwift19}),
but is clearly contaminated by the afterglow component (Fig.~\ref{fig:lc}) and does not provide a good measure of the actual duration of the prompt emission. 
A more detailed study of the prompt emission phase is reported in \cite{FermiSwift19}.

\begin{figure}
\centering
\includegraphics[width=0.8\textwidth]{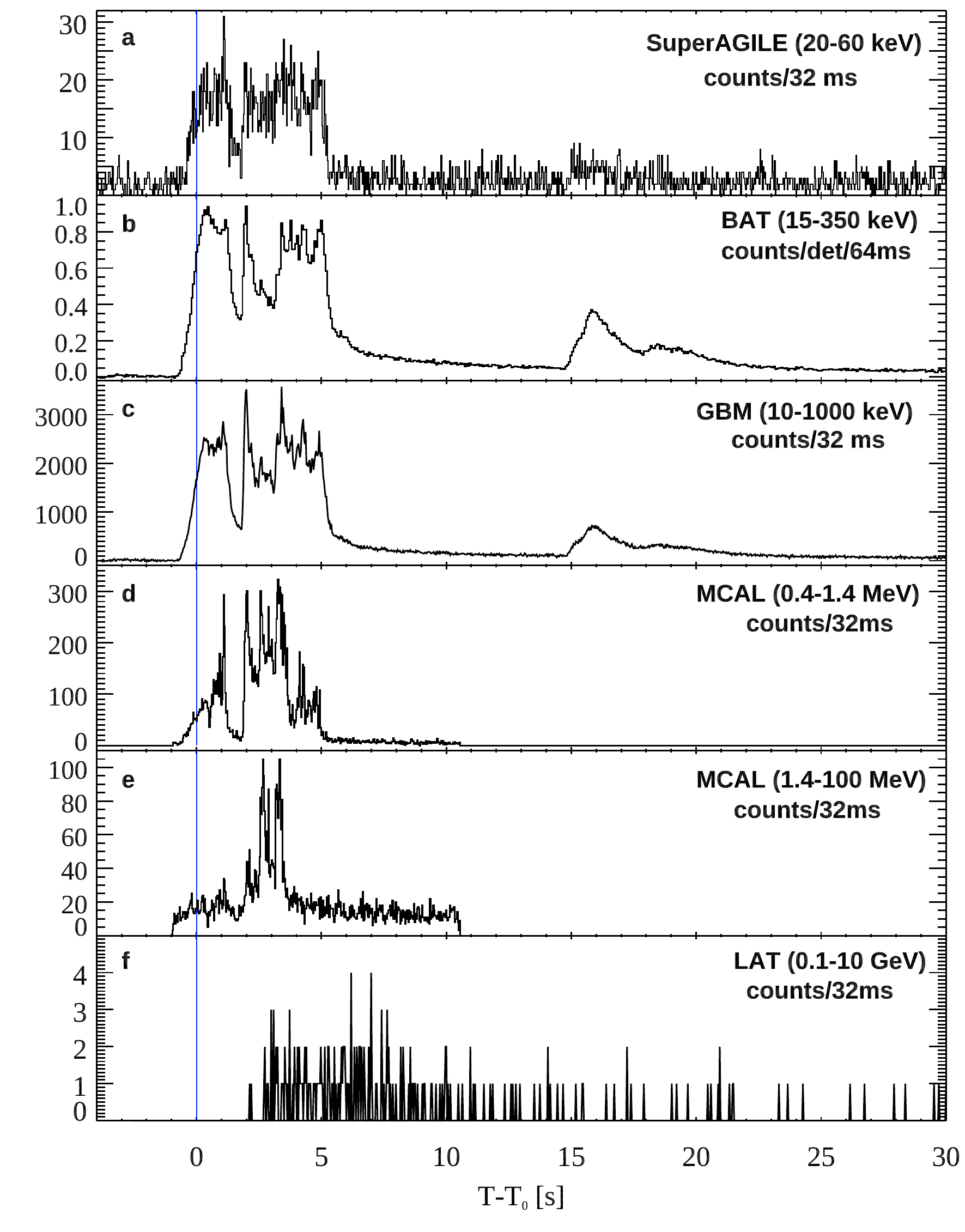}
\vskip -0.8truecm
\caption{{\bf Prompt emission light curves for different detectors.} The different panels show light curves for: {\bf a}, SuperAGILE (20-60\,keV); {\bf b}, {\it Swift}/BAT (15-150\,keV); {\bf c}, {\it Fermi}/GBM (10-1000\, keV); {\bf d}, {\it AGILE}/MCAL (0.4-1.4\,MeV); {\bf e}, {\it AGILE}/MCAL (1.4-100\,MeV); {\bf f}, {\it Fermi}/LAT (0.1-10\,GeV). The light curve of {\it AGILE}/MCAL is split into two bands to show the energy dependence of the first peak. Error bars show the 1-$\sigma$ statistical errors.}
\label{fig:prompt_lc}
\end{figure}

\subsection{AGILE} 
(The Astrorivelatore Gamma ad Immagini LEggero \cite{Tavani2009}) could observe GRB\,190114C until $T_0$+330\,s, before it became occulted by the Earth.
GRB\,190114C triggered the Mini-CALorimeter (MCAL) 
from T0$-$0.95\,s to T0$+$10.95\,s. 
The MCAL light flux curve in Fig.~\ref{fig:lc} has been produced using two different spectral models. 
From $T_0-$0.95\,s to $T_0+$1.8~s, the spectrum is fitted by a power law with photon index $\Gamma_{\rm ph}=$ -1.97$^{+0.47}_{-0.70}$ ($dN/dE\propto E^{\Gamma_{\rm ph}}$).
From $T_0+$1.8\,s to $T_0+5.5$\,s the best fit model is a broken power law with $\Gamma_{\rm ph,1}=-1.87^{+0.54}_{-0.19}$,  $\Gamma_{\rm ph,2}=-2.63^{+0.07}_{-0.07}$, and break energy $E_{\rm b}=$756$_{-159}^{+137}$\,keV.
The total fluence
in the 0.4$-$100\,MeV energy range is $F=1.75\times$10$^{-4}$\,erg\,cm$^{-2}$.
The Super-AGILE detector also detected the burst, but the large off-axis angle prevented any X-ray imaging of the burst, as well as spectral analysis.
Panels {\bf a}, {\bf d}, and {\bf e} in Extended Data Fig.~\ref{fig:prompt_lc} show the GRB\,190114C light curves acquired by the Super-AGILE detector ($20-60$\,keV) and by the MCAL detector in the low- ($0.4-1.4$\,MeV) and high-energy ($1.4-100$\,MeV) bands. 

\subsection{\textit{Fermi}/GBM}  
At the time of the MAGIC observations there are indications that some of the detectors are partially blocked by structure on the {\it Fermi} Spacecraft that is not modeled in the GBM detectors' response. This affects the low-energy part of the spectrum \cite{2012ApJS..199...19G}. For this reason, out of caution we elected to exclude the energy channels below 50\,keV.
The spectra detected by the {\it Fermi}-Gamma-ray Burst Monitor (GBM)\cite{Meegan+09GBM} during the $T_0$+68\,s to $T_0$+110\,s  and $T_0$+110\,s to $T_0$+180\,s intervals are best described by a power law model with photon index $\Gamma_{\rm ph}=-2.10\pm0.08$ and $\Gamma_{\rm ph}=-2.05\pm0.10$ respectively (Fig.~\ref{fig:SED_MAGIC_all} and Fig.~\ref{fig:SED_68-180}). 
The 10-1000\,keV light curve in Extended Data Fig.~\ref{fig:prompt_lc} (panel {\bf c}) was constructed by summing photon counts for the bright NaI detectors.

\subsection{\textit{Swift}/BAT}  
The $15-350$\,keV mask-weighted light curve of the Burst Alert Telescope (BAT \cite{2005SSRv..120..143B}) shows a multi-peaked structure that starts at $T_0-7$\,s (Extended Data Fig.~\ref{fig:prompt_lc}, panel {\bf b}).
The $68-110$\,s and $110-180$\,s spectra shown in Figs.~\ref{fig:SED_MAGIC_all} and \ref{fig:SED_68-180} were derived from joint XRT-BAT fit. 
The best-fitting parameters for the whole interval ($68-180$\,s) are: column density $N_{\rm H}=(7.53_{-1.74}^{+0.74})\times 10^{22}$\,cm$^{-2}$ at ${z=0.42}$, in addition to the galactic value of $7.5\times 10^{19}$\,cm$^{-2}$,
low-energy photon index $\Gamma_{\rm ph,1}=-1.21_{-1.26}^{+0.40}$, high-energy spectral index $\Gamma_{\rm ph,2}=-2.19_{-0.19}^{+0.39}$, peak energy $E_{\rm pk} > 14.5$\,keV.
Errors are given at $90\%$ confidence level.

\subsection{\textit{Fermi}/LAT}
The \textit{Fermi} Large Area Telescope (LAT)\cite{LATPaper} detected a gamma-ray counterpart since the prompt phase\cite{Kocevski2019GCN}. The burst left the LAT field of view (FoV) at $T_{\rm 0}$+150\,s and remained outside the LAT FoV until $T_{\rm 0}$+8600\,s.
The count light curve in the energy range 0.1-10\,GeV is shown in Extended Data Fig.~\ref{fig:prompt_lc} (panel {\bf f}).
The LAT spectra in the time bins 68--110\,s and 110--180\,s (Figs.~\ref{fig:SED_MAGIC_all} and \ref{fig:SED_68-180}) are described by a power law with pivot energies of, respectively, 200\,MeV and 500\,MeV, photon indices $\Gamma_{\rm ph} (68-110) = -2.02 \pm 0.95$ and $\Gamma_{\rm ph} (110-180) = -1.69 \pm 0.42$, and corresponding normalisations of $N_{0, 68-110} = (2.02 \pm 1.31) \times 10^{-7} ~ \textup{ph} ~ \textup{MeV}^{-1} \textup{cm}^{-2} ~ \textup{s}$ and $N_{0, 110-180} = (4.48 \pm 2.10) \times 10^{-8} ~ \textup{ph} ~ \textup{MeV}^{-1} \textup{cm}^{-2} ~ \textup{s}$. In each time-interval, the analysis has been performed limited to the energy range where photons have been detected.
The LAT light curve integrated in the energy range 0.1-1\,GeV is shown in Fig.~\ref{fig:lc}.

\subsection{MAGIC}
We used the Major Atmospheric Gamma Imaging Cherenkov (MAGIC) standard software \cite{Aleksicetal2016b} and followed the steps optimised for the data taking under moderate moon illumination\cite{Ahnen2017} to analyse the data.
The spectral fitting is performed by a forward-folding method assuming a simple power law for the intrinsic spectrum and taking into account the extragalactic background light (EBL) effect using the model of Dom{\'\i}nguez et al.\cite{Dominguezetal11}. 
Extended Data Table~\ref{tab:magicSfit} shows the fitting results for various time bins (the pivot energy is chosen to minimise the correlation between normalisation and photon index parameters). 
The data points shown in both Fig.~\ref{fig:SED_MAGIC_all} and \ref{fig:SED_68-180} are obtained from the observed excess rates in estimated energy whose fluxes are evaluated in true energy using effective time and a spill-over corrected effective area obtained as a resultant of the best fit.

\begin{table}
\centering
\begin{tabular}[th!]{cccc}
\toprule
Time bin & Normalisation  & Photon index & Pivot energy \\\relax
[\,seconds after $T_0$\,] & [\,\si{TeV^{-1}\,\cm^{-2}\,\second^{-1}}\,] & & [\si{GeV}] \\
\midrule
$62$ - $90$    & $\,1.95^{+0.21}_{-0.20}\,\cdot 10^{-7}$    & $-2.17\,^{+0.34}_{-0.36}\,$ & $395.5$ \\
\midrule
$68$ - $180$    & $\,1.10^{+0.09}_{-0.08}\,\cdot 10^{-7}$    & $-2.27\,^{+0.24}_{-0.25}\,$ & $404.7$ \\

$180$ - $625$    & $\,2.26^{+0.21}_{-0.20}\,\cdot 10^{-8}$    & $-2.56\,^{+0.27}_{-0.29}\,$ & $395.5$ \\
\midrule
$68$ - $110$    & $\,1.74^{+0.16}_{-0.15}\,\cdot 10^{-7}$    & $-2.16\,^{+0.29}_{-0.31}\,$ & $386.5$ \\

$110$ - $180$    & $\,8.59^{+0.95}_{-0.91}\,\cdot 10^{-8}$    & $-2.51\,^{+0.37}_{-0.41}\,$ & $395.5$ \\

$180$ - $360$    & $\,3.50^{+0.38}_{-0.36}\,\cdot 10^{-8}$    & $-2.36\,^{+0.34}_{-0.37}\,$ & $395.5$ \\

$360$ - $625$    & $\,1.65^{+0.23}_{-0.23}\,\cdot 10^{-8}$    & $-3.16\,^{+0.48}_{-0.54}\,$ & $369.1$ \\

$625$ - $2400$    & $\,3.52^{+0.47}_{-0.47}\,\cdot 10^{-9}$    & $-2.80\,^{+0.48}_{-0.54}\,$ & $369.1$ \\
\midrule
$62$ - $2400$ (Nominal MC)    & $\,1.07^{+0.08}_{-0.07}\,\cdot 10^{-8}$    & $-2.51\,^{+0.20}_{-0.21}\,$ & $423.8$ \\

$62$ - $2400$ (Light scale +15\% MC)    & $\,7.95^{+0.58}_{-0.56}\,\cdot 10^{-9}$    & $-2.91\,^{+0.23}_{-0.25}\,$ & $369.1$ \\

$62$ - $2400$ (Light scale -15\% MC)    & $\,1.34^{+0.09}_{-0.09}\,\cdot 10^{-8}$    & $-2.07\,^{+0.18}_{-0.19}\,$ & $509.5$ \\

\bottomrule
\end{tabular}
\caption{
{\bf MAGIC spectral fit parameters for GRB\,190114C.} 
For each time bin, columns represent a) start time and end time of the bin; b) normalisation of the EBL-corrected differential flux at the pivot energy with statistical errors; c) photon indices with statistical errors; d) pivot energy of the fit (fixed).}
\label{tab:magicSfit}
\end{table}

The time resolved analysis hints to a possible spectral evolution to softer values. Although we can not exclude that the photon indices are compatible with a constant value of $\sim -2.5$ up to 2400\,s.  
The signal and background in the considered time bins are both in the low-count Poisson regime. Therefore, 
the correct treatment of the MAGIC data provided here includes along with the use of the Poisson statistic also the systematic errors. 
To estimate the main source of systematic error caused by our imperfect knowledge of the absolute instrument calibration and the total atmospheric transmission we vary the light-scale in our Monte Carlo (MC) simulation as suggested in previous studies\cite{Aleksicetal2016b}. The result is reported in the last two lines of Extended Data Table~\ref{tab:magicSfit} and in Extended Data Fig.~\ref{fig:sederr}.

The systematic effects deriving from the choice of one particular EBL model were also studied. The analysis performed to obtain the time integrated spectrum was repeated employing other three models\cite{Franceschinietal2008,Finkeetal2010,Gilmoreetal2012}. The contribution to the systematic error on the photon index caused by the uncertainty on the EBL model is $\sigma_{\alpha} = ^{+0.10}_{-0.13}$ which is smaller than the statistical error only (1 standard deviation) as already seen in a previous work\cite{DiscoveryPaper}. On the other hand, the contribution to the systematic error on the normalisation, due to choice of the EBL model, is only partially at the same level of the statistical error (1 standard deviation) $\sigma_{N} = ^{+0.30}_{-0.08} \times 10^{-8}$. The chosen EBL model returns a lower normalisation with respect to two of the other models and very close to the third one~\cite{Franceschinietal2008}.

\begin{figure}
\centering
\includegraphics[width=0.9\textwidth]{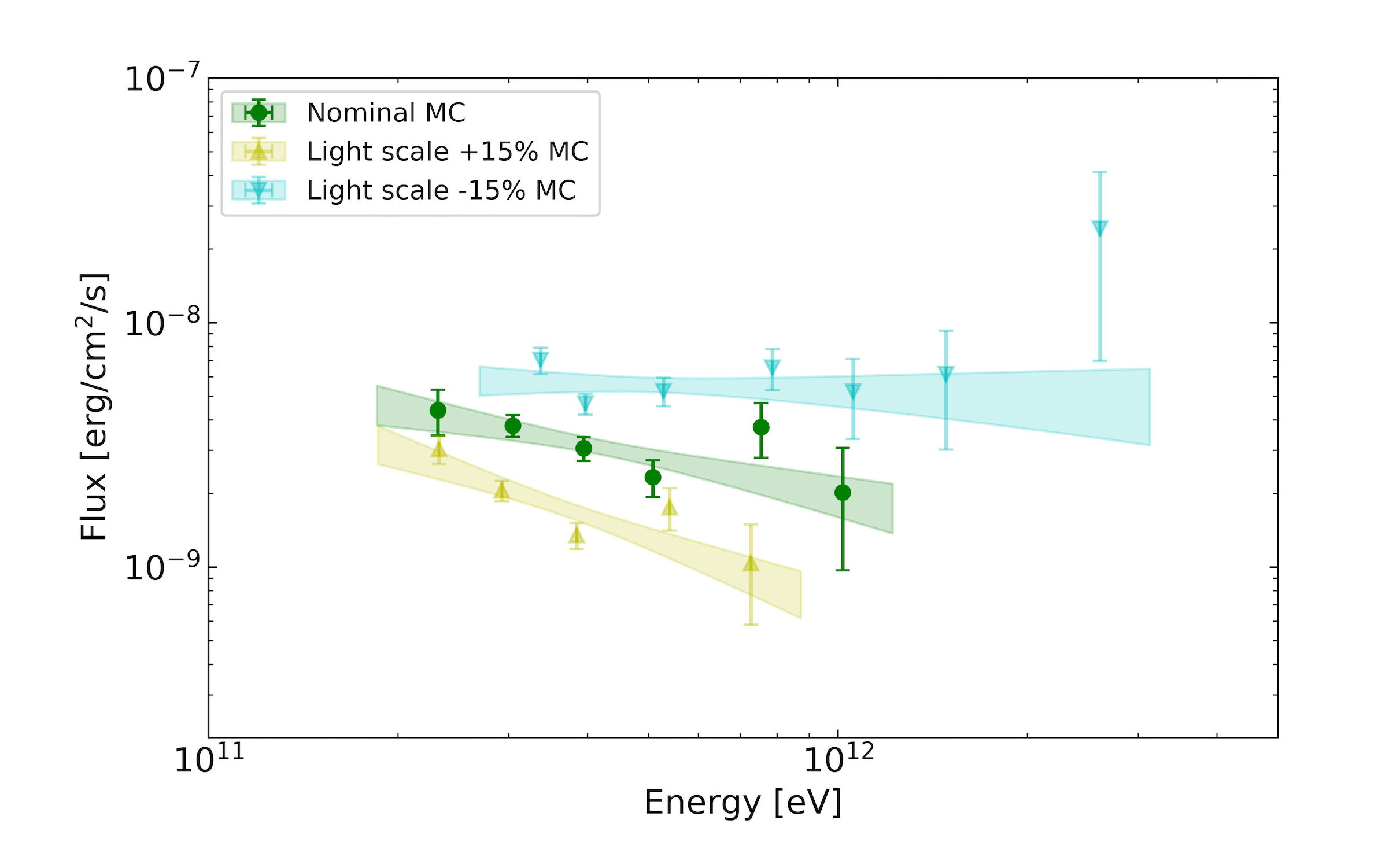}
\caption{
{\bf MAGIC time integrated spectral energy distributions in the time interval 62-2400\,s after $T_0 $.}
The green (yellow, blue) points and band show the result with the nominal (+15\%, -15\%) light scale MC, defining the limits of the systematic uncertainties. The contour regions are drawn from the 1-$\sigma$ error of their best-fit power law functions. The vertical bars of the data points show the 1-$\sigma$ errors on the flux.
}
\label{fig:sederr}
\end{figure}

The MAGIC energy flux light curve that is presented in Fig.~\ref{fig:lc} was obtained by integrating the best fit spectral model of each time bin from 0.3 to 1 TeV, in the same manner as a previous publication\cite{DiscoveryPaper}.
The value of the fitted time constant reported here differs less than two standard deviation from the one previously reported\cite{DiscoveryPaper}. The difference is due to the poor constraints on the spectral fit parameters of the last time bin, which influences the light curve fit.  

\newpage
\subsection{X-ray afterglow observations}
\subsection{\textit{Swift}/XRT}
The {\it Swift} X-Ray Telescope (XRT) 
started observing 68\,s after $T_0$. 
The source light curve\cite{UKXRTLC} was taken from the {\it Swift}/XRT light curve repository \cite{Evansetal2009} and converted into 1-10\,keV flux (Fig.~\ref{fig:lc}) through dedicated spectral fits.
The combined spectral fit XRT+BAT in Figs.~\ref{fig:SED_MAGIC_all} and \ref{fig:SED_68-180} has been described above.

\subsection{\textit{XMM-Newton} and \textit{NuSTAR}}
The \textit{XMM-Newton} X-ray Observatory 
and {\bf the} Nuclear Spectroscopic Telescope Array (\textit{NuSTAR})
started observing GRB\,190114C under DDT ToOs 
7.5\,hours and 22.5\,hrs (respectively) after the burst.
The \textit{XMM-Newton} and NuSTAR absorption-corrected fluxes (see Fig.~\ref{fig:lc}) were derived by fitting the spectrum with XSPEC adopting the same power law model, with absorption in our Galaxy and at the redshift of the burst.

\subsection{NIR, Optical and UV afterglow observations} Light curves from the different instruments presented in this section are shown in Extended Data Fig.~\ref{fig:optical_lc}.

\begin{figure}
\centering
\includegraphics[width=0.5\textwidth]{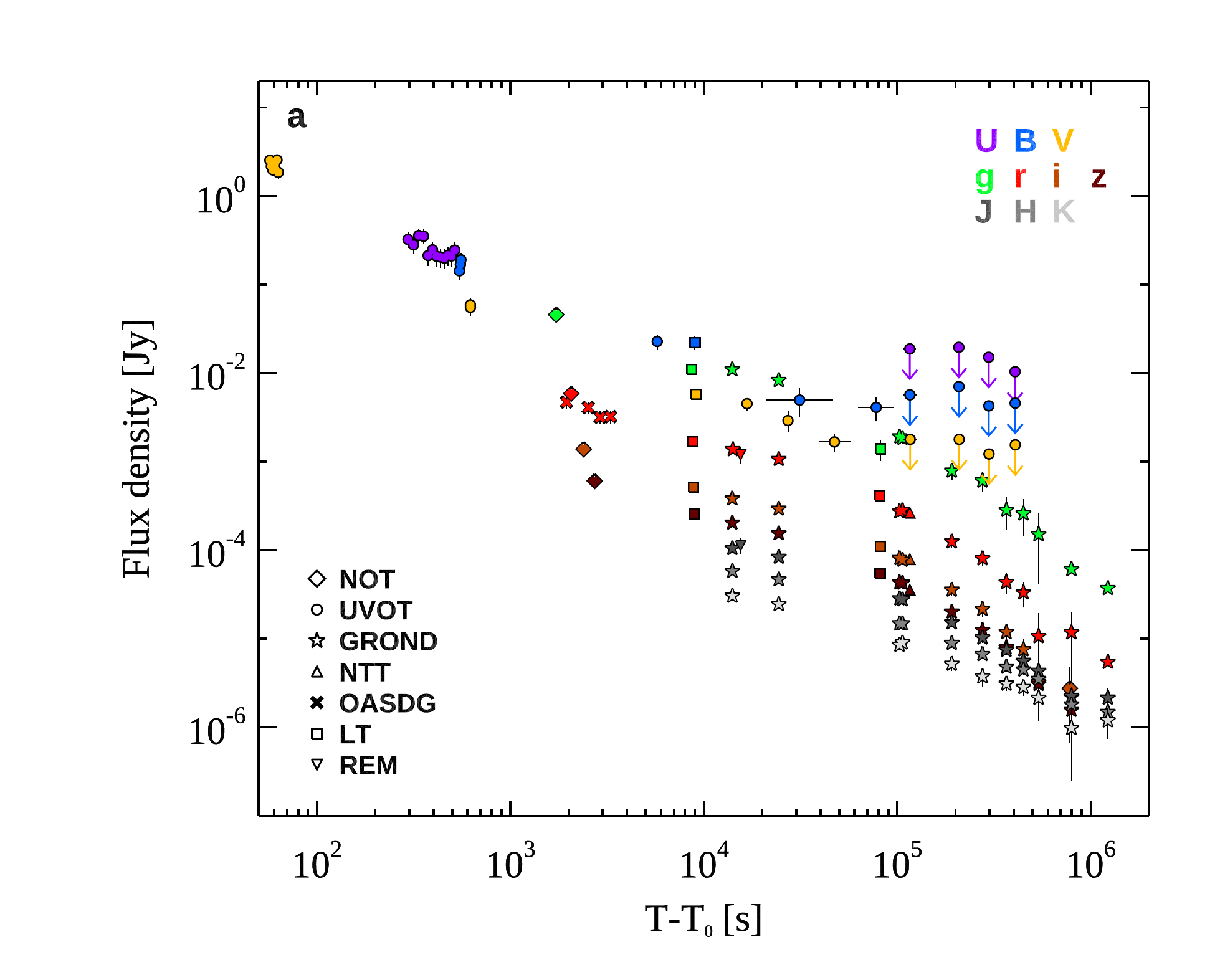}\includegraphics[width=0.5\textwidth]{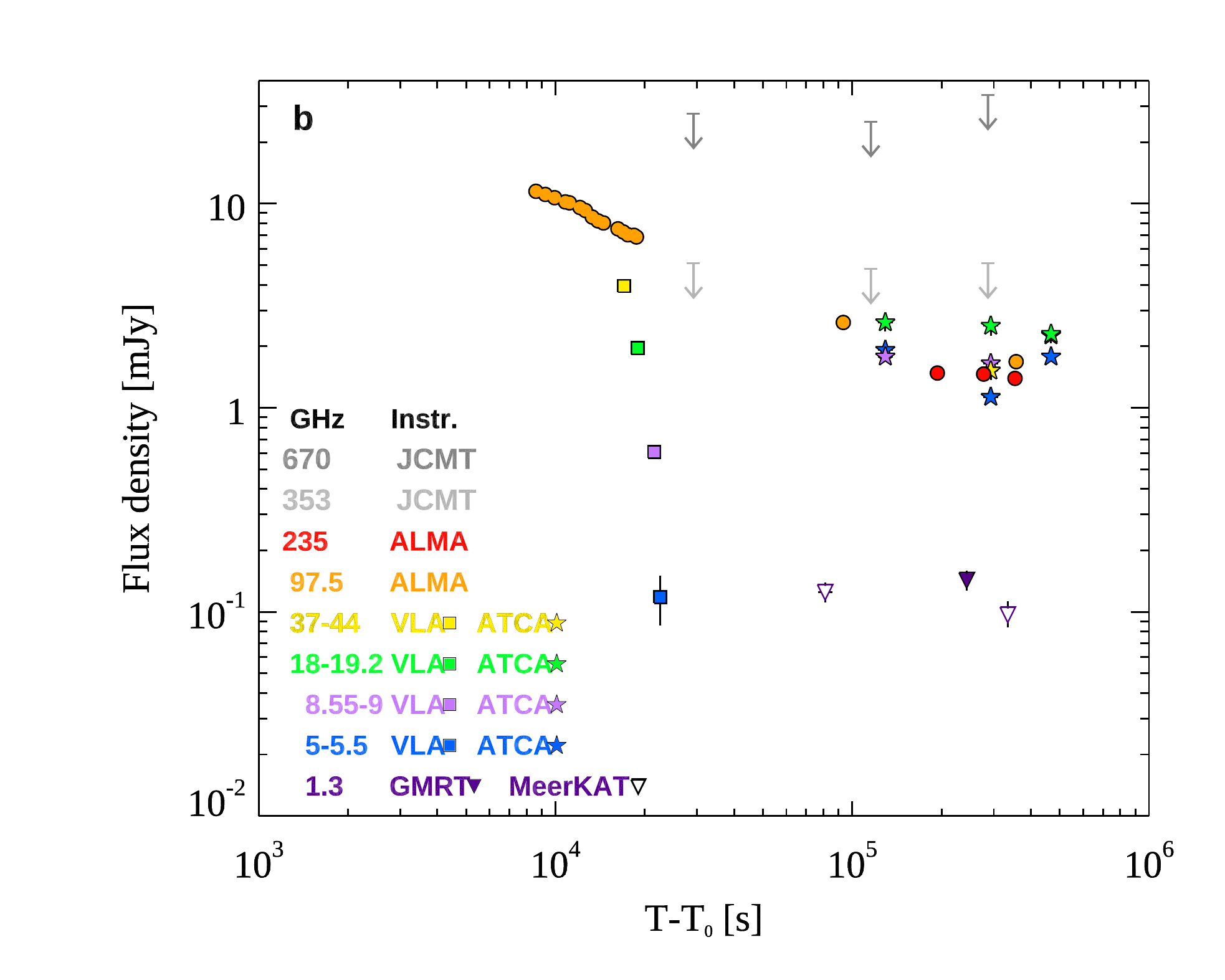}
\caption{ {\bf Afterglow light curves of GRB~190114C.} Flux density at different frequencies, as a function of the time since the initial burst $T_0$. Panel {\bf a}: observations in the NIR/Optical/UV bands.
The flux has been corrected for extinction in the host and in our Galaxy. The contribution of the host galaxy and its companion has been subtracted. 
Fluxes have been rescaled (except for the $r$ filter).
Panel {\bf b}:  Radio and sub-mm observations from 1.3\,GHz to 670\,GHz.}
\label{fig:radio_lc}
\label{fig:optical_lc}
\end{figure}

\subsection{GROND} The Gamma-ray Burst Optical/Near-infrared Detector (GROND\cite{Greineretal2008}) 
started observations 3.8 hours after the GRB trigger, and the follow-up continued until January 29, 2019. 
Image reduction and photometry were carried out with standard IRAF tasks \cite{Tody1993},
as described in \cite{Kruhleretal2008,Bolmeretal2018}.
\JHK\ photometry was converted to AB magnitudes to have a common flux system. 
Final photometry is given in Extended Data Table~\ref{tab:grondphot}.

\begin{table}
\tiny
    \centering
    \begin{tabular}{cccccccc}
\hline\hline
\multicolumn{1}{c}{$T_{\mathrm{GROND}}$}   & \multicolumn{7}{c}{AB magnitude} \\
\multicolumn{1}{c}{(s)}       & $g'$  & $r'$ & $i'$ & $z'$ & $J$ & $H$ & $K_s$ \\ 
\hline
$14029.94\pm 335.28$    &  $19.21\pm 0.03$ & $18.46\pm 0.03$ & $17.78\pm 0.03$ & $17.33\pm 0.03$ & $16.78\pm 0.05$ & $16.30\pm 0.05$ & $16.03\pm 0.07$ \\
$24402.00\pm 345.66$    &  $19.50\pm 0.04$ & $18.72\pm 0.03$ & $18.05\pm 0.03$ & $17.61\pm 0.03$ & $17.02\pm 0.05$ & $16.53\pm 0.05$ & $16.26\pm 0.08$ \\
$102697.17\pm 524.01$   &  $20.83\pm 0.06$ & $20.00\pm 0.04$ & $19.30\pm 0.04$ & $18.87\pm 0.03$ & $18.15\pm 0.05$ & $17.75\pm 0.06$ & $17.40\pm 0.09$ \\
$106405.63\pm 519.87$   &  $20.86\pm 0.05$ & $19.98\pm 0.03$ & $19.34\pm 0.03$ & $18.88\pm 0.03$ & $18.17\pm 0.06$ & $17.75\pm 0.06$ & $17.34\pm 0.09$ \\
$191466.77\pm 751.37$   &  $21.43\pm 0.07$ & $20.61\pm 0.03$ & $19.97\pm 0.03$ & $19.52\pm 0.03$ & $18.77\pm 0.06$ & $18.28\pm 0.06$ & $17.92\pm 0.14$ \\
$275594.19\pm 747.59$   &  $21.57\pm 0.07$ & $20.88\pm 0.04$ & $20.31\pm 0.04$ & $19.87\pm 0.04$ & $19.14\pm 0.07$ & $18.57\pm 0.06$ & $18.26\pm 0.21$ \\
$366390.74\pm 1105.79$  &  $21.87\pm 0.07$ & $21.17\pm 0.04$ & $20.62\pm 0.03$ & $20.15\pm 0.03$ & $19.43\pm 0.06$ & $18.89\pm 0.06$ & $18.46\pm 0.15$ \\
$448791.55\pm 1201.33$  &  $21.90\pm 0.08$ & $21.27\pm 0.04$ & $20.79\pm 0.04$ & $20.33\pm 0.03$ & $19.66\pm 0.07$ & $18.97\pm 0.07$ & $18.55\pm 0.18$ \\
$537481.41\pm 1132.16$  &  $22.02\pm 0.09$ & $21.52\pm 0.05$ & $21.00\pm 0.04$ & $20.55\pm 0.03$ & $19.87\pm 0.07$ & $19.20\pm 0.07$ & $18.83\pm 0.17$ \\
$794992.63\pm 1200.69$  &  $22.14\pm 0.04$ & $21.51\pm 0.03$ & $21.05\pm 0.04$ & $20.71\pm 0.05$ & $20.31\pm 0.13$ & $19.79\pm 0.14$ & $19.59\pm 0.41$ \\
$1226716.84\pm 1050.15$ &  $22.17\pm 0.04$ & $21.59\pm 0.04$ & $21.26\pm 0.04$ & $20.97\pm 0.04$ & $20.34\pm 0.12$ & $19.95\pm 0.11$ & $19.40\pm 0.34$ \\ \hline
    \end{tabular}
    \caption{{\bf GROND photometry.} $T_{\mathrm{GROND}}$ in seconds after the BAT trigger. The AB magnitudes are not
    corrected for the Galactic foreground reddening.}
    \label{tab:grondphot}
\end{table}

\subsection{GTC}
The BOOTES-2 ultra-wide field camera \cite{Castro-Tiradoetal2008}, took an image at the GRB 190114C location, starting at 20:57:18 UT (30\,s exposure time) (see Extended Data Fig.~\ref{fig:allskyJan14F}).
The Gran Canarias Telescope (GTC) equipped with the OSIRIS spectrograph\cite{Cepaetal2000}  started observations 2.6\,hr post-burst.
The grisms R1000B and R2500I were used covering the wavelength range 3,700-10,000\,\AA\ (600\,s exposure times for each grism).
The GTC detects a highly extinguished continuum, as well as CaII H and K lines in absorption, and [OII], H$_{\beta}$, and [OIII] in emission (see Extended Data Fig.~\ref{fig:GTCspectrum}), all roughly at the same redshift $z$ = 0.4245$\pm$0.0005 \cite{Castro-Tirado2019}.
Comparing the derived rest-frame equivalent widths (EWs) with the work by \cite{deUgartePostigoetal2012}, GRB\,190114C clearly shows higher than average, but not unprecedented, values.

\subsection{HST} The {\it Hubble Space Telescope} ({\it HST}) imaged the afterglow and host galaxy of GRB\,190114C on 11 February and 12 March 2019. 
HST observations clearly reveal that the host galaxy is spiral (Extended Data Fig.~\ref{fig:3colHST}).
A direct subtraction of the epochs of F850LP observations yields a faint residual close to the nucleus of the host (Extended Data Fig.\,\ref{fig:diffHST}). 
From the position of the residual we estimate that the burst originated within 250\,pc of the host galaxy nucleus.

\begin{figure}
\centering
\includegraphics{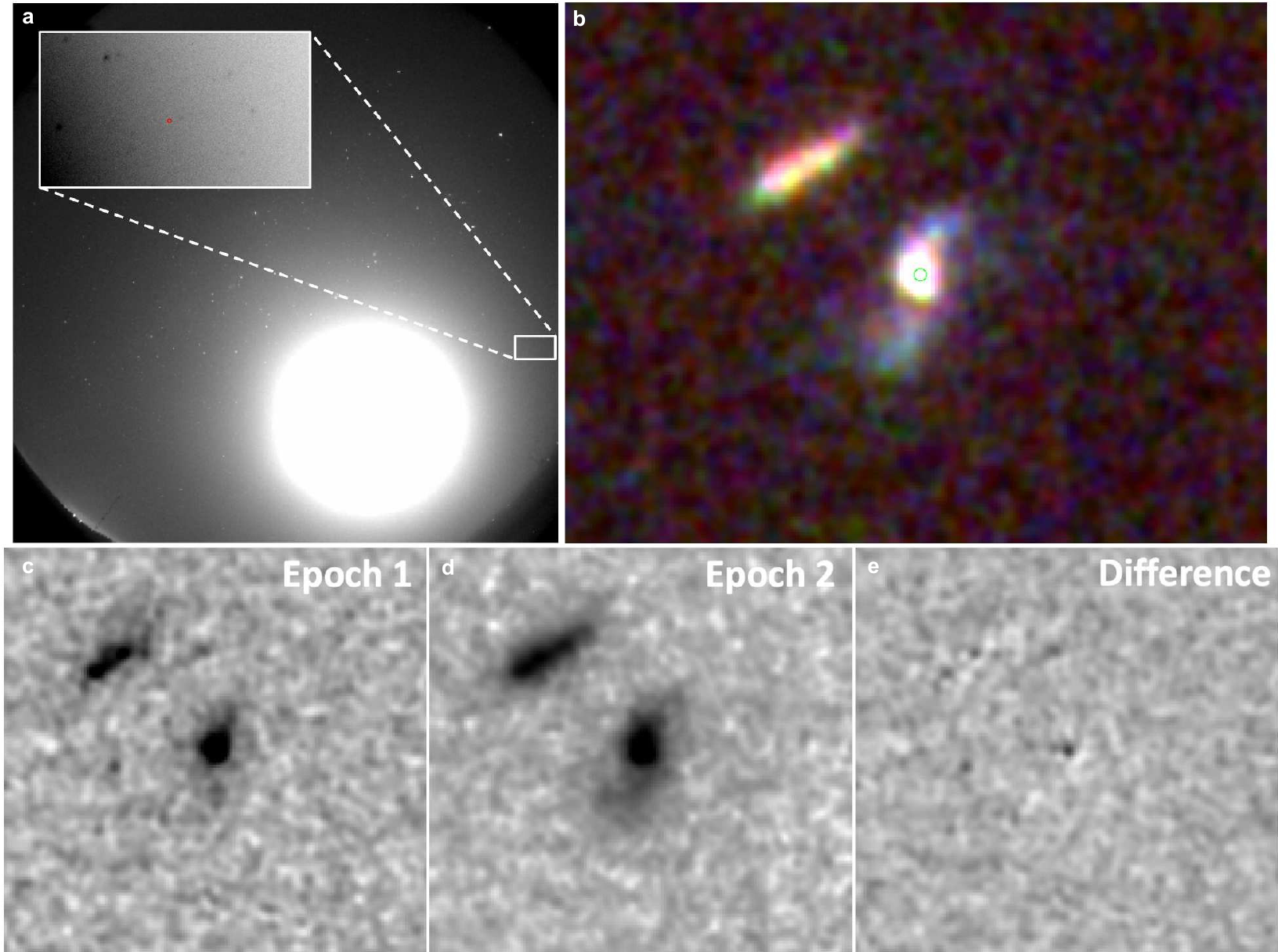}

\caption{{\bf Images of the localisation region of GRB~190114C.} Panel {\bf a}: The CASANDRA-2 at the BOOTES-2 station all-sky image. The image (30\,s exposure, unfiltered) was taken at T0+14.8\,s. At the GRB\,190114C location (circle) no prompt optical emission is detected.
Panel {\bf b}: Three-colour image of the host of GRB\,190114C with the HST. The host galaxy is a spiral galaxy, and the green circle indicates the location of the transient close to its host nucleus. The image is $8$\,\arcsec\  across, north is up and east to the left.
Panels {\bf c, d} and {\bf e}: 
F850LP imaging of GRB\,190114C taken with the HST. Two epochs are shown (images are 4\,\arcsec\  across), as well as the result of the difference image. A faint transient is visible close to the nucleus of the galaxy, and we identify this as the late time afterglow of the burst. 
}
\label{fig:diffHST}
\label{fig:3colHST}
\label{fig:allskyJan14F}
\end{figure}

\subsection{LT} The robotic 2-m Liverpool Telescope (LT\cite{Steeleetal2004})
slewed to the afterglow location at UTC 2019-01-14.974 and on the second night, from UTC 2019-01-15.814 and acquired images in $B$, $g$, $V$, $r$, $i$ and $z$ bands (45\,s exposure each in the first night and 60\,s in the second, see Extended Data Table\,\ref{tab:lt_table}).  
Aperture photometry of the afterglow was performed using a custom IDL script with a fixed aperture radius of 1.5\arcsec. Photometric calibration was performed relative to stars from the Pan-STARRS1 catalogue\cite{Chambersetal2016}.

\begin{table}
\tiny
       \centering
       \renewcommand{\arraystretch}{0.6}
       \begin{tabular}{cccc}
               \hline
               UTC   & Filter & Exposure (s) & Magnitude\\
               \hline
               \multicolumn{4}{c}{LT/IO:O}\\
               2019-01-14.975  & $g$ & 45 & 19.08$\pm$0.06\\
               2019-01-14.976  & $r$ & 45 & 18.22$\pm$0.02\\
               2019-01-14.977  & $i$ & 45 & 17.49$\pm$0.02\\
               2019-01-14.978  & $z$ & 45 & 17.12$\pm$0.02\\
               2019-01-14.979  & $B$ & 45 & 19.55$\pm$0.15\\
               2019-01-14.980  & $V$ & 45 & 18.81$\pm$0.08\\
               2019-01-15.814  & $r$ & 60 & 19.61$\pm$0.05\\
               2019-01-15.818  & $z$ & 60 & 18.70$\pm$0.06\\
               2019-01-15.820  & $i$ & 60 & 19.04$\pm$0.04\\
               2019-01-15.823  & $g$ & 60 & 20.96$\pm$0.17\\
               \hline
               \hline
                       \multicolumn{4}{c}{NOT/AlFOSC}\\
               2019-01-14.89127 &  $g$ & $1 \times 300$ & 17.72$\pm$0.03\\
               2019-01-14.89512 &  $r$ & $1 \times 300$ & 16.93$\pm$0.02\\
               2019-01-14.89899 &  $i$ & $1 \times 300$ & 16.42 $\pm$0.04\\
               2019-01-14.90286 &  $z$ & $1 \times 300$ & 16.17 $\pm$0.04\\
               2019-01-23.8896  &  $i$ & $6 \times 300$ & 21.02$\pm$0.05\\
    \hline
    \end{tabular}
\begin{tabular}{cccc|cccc}
               \toprule
               \multicolumn{8}{c}{UVOT}\\
               \midrule
    $T_{\rm start}$  & $T_{\rm stop}$ & Filter  & Magnitude & $T_{\rm start}$  & $T_{\rm stop}$ & Filter  & Magnitude\\
	\midrule
    56.63  &   57.63  &  $V    $ & 12.17$\pm$0.14 & 130958  &   142524  &   $UVM2 $    &    20.37      \\
    57.63  &   58.63  &  $V    $ & 12.34$\pm$0.14 &   217406  &   222752  &   $UVM2 $    &    20.48      \\
    58.63  &   59.63  &  $V    $ & 12.44$\pm$0.13 &   107573  &   125233  &   $U  $    &    20.29      \\
    59.63  &   60.63  &  $V    $ & 12.29$\pm$0.14 &   205500  &   210750  &   $U  $    &    20.25      \\
    60.63  &   61.63  &  $V    $ & 12.44$\pm$0.14 &   291188  &   302718  &   $U  $    &    20.49      \\
    61.63  &   62.63  &  $V    $ & 12.16$\pm$0.13 &   400429  &   412385  &   $U  $    &    20.82      \\
    62.63  &   63.63  &  $V    $ & 12.51$\pm$0.13 &      616  &  627  &   $V   $    &    16.25$\pm$0.20   \\  
   615.95  &  625.95  &  $V    $ & 16.32$\pm$0.20 &    16295  &    17136  &   $V  $    &    19.03$\pm$0.14   \\  
    73.34  &   83.34  &  $white$  & 13.86    &    26775  &    27682  &   $V  $    &    19.50$\pm$0.27   \\  
    83.34  &   93.34  &  $white$  & 14.10$\pm$0.06 &    39149  &    57221  &   $V  $    &    20.09$\pm$0.23   \\  
    93.34  &   103.34 &  $white$  & 14.19$\pm$0.06 &   108064  &   125736  &   $V  $    &    20.02            \\
   103.34  &  113.34  &  $white$  & 14.36$\pm$0.06 &   206689  &   211356  &   $V  $    &    20.02            \\
   113.34  &  123.34  &  $white$  & 14.64$\pm$0.06 &   292383  &   303996  &   $V  $    &    20.42            \\
   123.34  &  133.34  &  $white$  & 14.65$\pm$0.06 &   401305  &   413316  &   $V  $    &    20.17           \\
   133.34  &  143.34  &  $white$  & 14.91$\pm$0.06 &     4044  &    51522  &   $UVW1 $    &    21.17           \\
   143.34  &  153.34  &  $white$  & 14.99$\pm$0.06 &   131216  &   142656  &   $UVW1 $    &    20.47           \\
   153.34  &  163.34  &  $white$  & 15.05$\pm$0.06 &   217984  &   223056  &   $UVW1 $    &    20.57           \\
   163.34  &  173.34  &  $white$  & 15.32$\pm$0.06 &      592  &  612  &   $UVW2 $    &    17.65           \\
   173.34  &  183.34  &  $white$  & 15.38$\pm$0.06 &     6056  &    56384  &   $UVW2 $    &    21.30           \\
   183.34  &  193.34  &  $white$  & 15.38$\pm$0.06 &   130699  &   142346  &   $UVW2 $    &    20.52           \\
   193.34  &  203.34  &  $white$  & 15.59$\pm$0.06 &   216828  &   222404  &   $UVW2 $    &    20.55           \\
   562.0   &  572.0   &  $white$  & 16.96$\pm$0.10 &      566  &  586  &   $white$    &    16.90$\pm$0.07  \\  
   572.0   &  582.0   &  $white$  & 16.90$\pm$0.10 &   607389  &   613956  &   $white$    &    22.16            \\
   535.5   &  555.5   &  $B    $  & 17.56$\pm$0.21 &   624452  &   682416  &   $white$    &    21.99$\pm$0.18  \\    
   545.5   &  565.5   &  $B    $  & 17.25$\pm$0.18 &   745033  &   769296  &   $white$    &    21.64$\pm$0.16  \\     
   285.9   &  305.9   &  $U    $  & 17.35$\pm$0.19 &   818840  &   837216  &   $white$    &    22.50      \\
   305.9   &  325.9   &  $U    $  & 17.50$\pm$0.20 &   893522  &   907116  &   $white$    &    22.57      \\
   325.9   &  345.9   &  $U    $  & 17.24$\pm$0.18 &   991065  &  1004196  &   $white$    &    22.49$\pm$0.35  \\    
   345.9   &  365.9   &  $U    $  & 17.26$\pm$0.18 &  1077542  &  1094616  &   $white$    &    22.41$\pm$0.31  \\   
   365.9   &  385.9   &  $U    $  & 17.80$\pm$0.24 &  1140343  &  1170336  &   $white$    &    22.50            \\
   385.9   &  405.9   &  $U    $  & 17.64$\pm$0.22 &  1220661  &  1274376  &   $white$    &    22.36$\pm$0.29  \\    
   405.9   &  425.9   &  $U    $  & 17.82$\pm$0.24 &     5851  & 6050  &   $white$    &    19.25$\pm$0.09  \\  
   425.9   &  445.9   &  $U    $  & 17.84$\pm$0.25 &    21950  &    22857  &   $white$    &    20.25$\pm$0.09  \\    
   445.9   &  465.9   &  $U    $  & 17.87$\pm$0.25 &  1353459  &  1359284  &   $white$    &    21.70            \\
   465.9   &  485.9   &  $U    $  & 17.79$\pm$0.24 &  1502211  &  1548336  &   $white$    &    21.98$\pm$0.24  \\    
   485.9   &  505.9   &  $U    $  & 17.81$\pm$0.24 &  1692292  &  1703935  &   $white$    &    22.07      \\
   505.9   &  525.9   &  $U    $  & 17.65$\pm$0.22 &  2132978  &  2146056  &   $white$    &    22.58      \\
     542  &  561  &   $B   $    &    17.38$\pm$0.14  &  2299521  &  2317956  &   $white$    &    22.41$\pm$0.31  \\    
    5646  & 5845  &   $B   $    &    19.54$\pm$0.19  &    63686  &    80942  &   $white$    &    21.07$\pm$0.24  \\    
   21038  &    46521  &   $B  $    &    21.14$\pm$0.35  &   107900  &   125591  &   $white$    &    21.40$\pm$0.28  \\    
   62774  &    96486  &   $B  $    &    21.33$\pm$0.29  &   206292  &   211137  &   $white$    &    21.52            \\
  107737  &   125412  &   $B  $    &    21.00           &   291984  &   303556  &   $white$    &    21.48$\pm$0.23  \\    
  205896  &   210944  &   $B  $    &    20.78           &   401012  &   413029  &   $white$    &    21.84            \\
  291586  &   303137  &   $B  $    &    21.29           &   491973  &   505356  &   $white$    &    22.21$\pm$0.24  \\    
  400721  &   412707  &   $B  $    &    21.22           &       74  &  224  &   $white$    &    14.90$\pm$0.02  \\  
    3839  &    50615  &   $UVM2 $    &    20.88$\pm$0.28 &         &   &       &      \\
\bottomrule
\end{tabular}
       \caption{{\bf Liverpool Telescope, Nordic Optical Telescope, and UVOT observations.} Magnitudes are SDSS AB-''like'' for ugriz, Vega-"like" for all the other filters and are not corrected for Galactic extinction. For UVOT data, magnitudes without uncertainties are upper limits.}
       \label{tab:lt_table}
\end{table} 

\subsection{NTT} The ESO New Technology Telescope (NTT) 
observed the optical counterpart of GRB\,190114C under the extended Public ESO Spectroscopic Survey for Transient Objects (ePESSTO) using the NTT/EFOSC2 instrument in imaging mode \cite{Tarenghi&Wilson1989}. 
Observations started at 04:36:53 UT on 2019 January 16 with the $g$, $r$, $i$, $z$ Gunn filters.
Image reduction was carried out by following the standard procedures\cite{Smarttetal2015}.

\subsection{OASDG} The 0.5\,m remote telescope of the Osservatorio Astronomico ``S. Di Giacomo" (OASDG), located in Agerola (Italy) started observations in the optical $Rc$-band 0.54 hours after the burst.
The afterglow of GRB\,190114C was clearly detected in all the images.

\subsection{NOT} The Nordic Optical Telescope (NOT) observed the optical afterglow of GRB\,190114C with the Alhambra Faint Object Spectrograph and Camera (AlFOSC) instrument. Imaging was obtained in the $griz$ filters with 300\,s exposures, starting at Jan 14 21:20:56 UT, 24 minutes after the BAT trigger.
The normalised spectrum (Extended Data Fig.~\ref{fig:NOTspectrum}) reveals strong host interstellar absorption lines due to Ca H \& K and Na I\,D, which provided a redshift of $z = 0.425$.

\begin{figure}
\centering
\includegraphics[scale=0.55,angle=0]{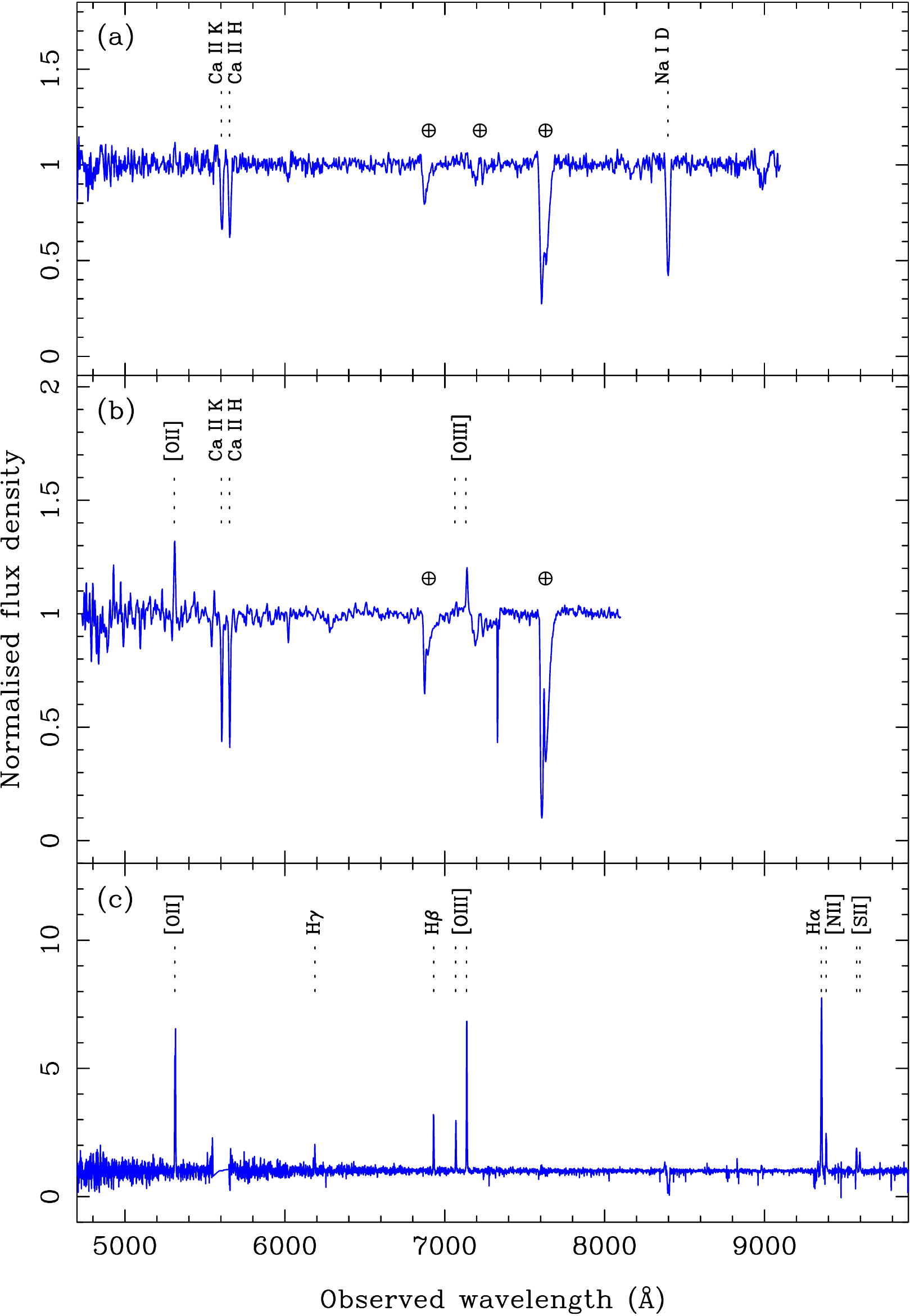}
\caption{{\bf Optical/NIR spectra of GRB~190114C.} 
Panel {\bf a}: The NOT/AlFOSC spectrum obtained at a mid-time 1\,hr post-burst. The continuum is afterglow dominated at this time, and shows 
strong absorption features of Ca\,II and Na\,I (in addition to telluric absorption).
Panel {\bf b}: the normalised  GTC (+OSIRIS) spectrum on Jan 14, 23:32:03 UT, with the R1000B and R2500I grisms. The emission lines of the underlying host galaxy are noticeable, besides the Ca II absorption lines in the afterglow spectrum. 
Panel {\bf c}: The visible light region of the VLT/X-shooter spectrum obtained approximately 3.2\,d post-burst, showing strong emission lines from the star-forming host galaxy.}
\label{fig:NOTspectrum}
\label{fig:GTCspectrum}
\label{fig:VLTspectrum}
\end{figure}

\subsection{REM} The Rapid Eye Mount telescope (REM) performed optical and NIR observations with the REM 60\,cm robotic telescope equipped with the ROS2 optical imager and the REMIR NIR camera\cite{Covinoetal2004}. Observations were performed 
starting 
about 3.8\,hours after the burst in the $r,$ and $J$ bands and lasted about one hour. 

\subsection{{\textit Swift}/UVOT}
The {\it Swift} UltraViolet and Optical Telescope (UVOT\cite{Roming2005}) began observations at $T_0+54$ seconds  in the UVOT $v$ band.
The first observation after settling started 74\,s after the trigger for 150\,s in the UVOT $white$ band\cite{2019Siegel}. 
A 50\,s exposure with the UV grism was taken thereafter, followed by multiple exposures rotating through all seven broad and intermediate-band filters until switching to only UVOT's clear white filter on 2019-01-20. 
Standard photometric calibration and methods were used for deriving the aperture photometry\cite{Poole2008,Breeveld2011}. 
The grism zeroth order the data were reduced manually\cite{Kuin2015} to derive the $b$-magnitude and error.

\subsection{VLT} The STARGATE collaboration used the Very Large Telescope (VLT)  and observed GRB\,190114C using the X-shooter spectrograph.
Detailed analysis will be presented in forthcoming papers.
A portion of the second spectrum is shown in Extended Data Fig.~\ref{fig:VLTspectrum}, illustrating the strong emission
lines characteristic of a strongly star-forming galaxy, whose light is largely dominating over the afterglow at this epoch.

\subsection{Magnitudes of the underlying galaxies}

The {\it HST} 
images show a spiral or tidally disrupted galaxy whose bulge is coincident with the coordinates of GRB\,190114C. A second galaxy is detected at an angular distance of 1.3$^{\prime\prime}$, towards the North East.
The SED analysis was performed with LePhare \cite{Arnouts1999,Ilbert2006} using an iterative method that combined both the resolved photometry of the two galaxies found in the {\it HST} and {\it VLT}/HAWK-I data and the blended photometry from {\it GALEX} and WISE, where the spatial resolution was much lower. Further details will be given in a paper in preparation (de Ugate Postigo et al.). The estimated photometry, for each object and their combination, is given in Extended Data Table~\ref{Table:host}.

\begin{table}
    \centering
\begin{tabular}{lccc}
\hline
   Filter        &        Host      &       Companion   &  Combined \\
           \hline
Sloan $u$   &    23.54	 &      25.74  &    23.40 \\
Sloan $g$  &    22.51	 &	23.81	&   22.21 \\
Sloan $r$  &    22.13	 &	22.81	&   21.66 \\
Sloan $i$  &    21.70	 &	22.27	&   21.19 \\
Sloan $z$  &    21.51	 &	21.74	&   20.87 \\
2MASS $J$  &    20.98	 &	21.08	&   20.28 \\
2MASS $H$  &    20.68	 &	20.82	&   20.00 \\
2MASS $Ks$ &    20.45	 &	20.61	&   19.77 \\
\hline
    \end{tabular}
    \caption{{\bf Observations of the host galaxy.} For each filter, the estimated magnitudes are given for the host galaxy of GRB~190114C, the companion and the combination of the two objects.}
    \label{Table:host}
\end{table}
 

\subsection{Optical Extinction}

The optical extinction toward the line of sight of a GRB
is derived assuming {\bf a} power law as intrinsic spectral shape\cite{Covinoetal2013}.
 Once the Galactic extinction ($E_{\rm B-V} = 0.01$\cite{Schlaflyetal2011}) is taken into account and the fairly bright host galaxy contribution is properly subtracted, a good fit to the data is obtained with the LMC recipe and $A_V = 1.83\pm0.15$. 
The spectral index $\beta$ ($F_\nu\propto\nu^{\beta_{\rm o}}$) evolves from hard to soft across the temporal break in the optical light-curve at about 0.5\,days, moving from $\beta_{\rm o,1}=-0.10\pm0.12$ to $\beta_{\rm o,2}=-0.48\pm0.15$. 

\subsection{Radio and Sub-mm afterglow observations} The light curves from the different instruments are shown in Extended Data Fig.~\ref{fig:radio_lc}.

\subsection{ALMA} The Atacama Large Millimetre/Submillimetre Array (ALMA) observations are reported in Band 3 (central observed frequency of 97.500\,GHz) and Band 6 (235.0487\,GHz), 
between 2019 January 15 and 2019 January 19.
Data were calibrated within CASA (Common Astronomy Software Applications, version 5.4.0\cite{McMullinetal2007}) using the pipeline calibration.
Photometric measurements were also performed within CASA. 
ALMA early observations at 97.5\,GHz are taken from \cite{Laskaretal2019}.

\subsection{ATCA} The Australia Telescope Compact Array (ATCA) observations were made with the ATCA 4\,cm receivers (band centres 5.5 and 9\,GHz), 15\,mm receivers (band centres 17 and 19\,GHz), and 7\,mm receivers (band centres 43 and 45\,GHz). 
ATCA data were obtained using the CABB continuum mode \cite{Wilsonetal2011a} and  reduced with the software packages {\sc Miriad} \cite{Saultetal1995a} and {\sc CASA} \cite{McMullinetal2007}
using standard techniques.
The quoted errors are $1\sigma$, which include the RMS and Gaussian $1\sigma$ errors.

\begin{table}
\footnotesize
    \centering
    \begin{tabular}{cccc}
    \toprule
    \multicolumn{4}{c}{ATCA} \\
    \midrule
    \vspace{-0.4 truecm}
         Start Date and Time &  End Date and Time   &    Frequency   & Flux \\
                    &                      &    GHz         & mJy  \\
\hline
1/16/2019 6:47:00   &  1/16/2019 10:53:00  &    5.5	    & 1.92$\pm$0.06\\
 	  	    &  		 	   &	9	    & 1.78$\pm$0.06\\
		    &			   &	18	    & 2.62$\pm$0.26\\
\hline
1/18/2019 1:45:00   & 1/18/2019 11:18:00   & 	5.5	    & 1.13$\pm$0.04\\
	  	    & 			   &	9	    & 1.65$\pm$0.05\\
		    &			   &	18	    & 2.52$\pm$0.27\\
		    &			   &	44	    & 1.52$\pm$0.15\\
\hline
1/20/2019 3:38	    & 1/20/2019 10:25:00   &	5.5	    & 1.78$\pm$0.06\\
	  	    & 			   &	9	    & 2.26$\pm$0.07\\
		    &			   &	18	    & 2.30$\pm$0.23\\
\hline
    \end{tabular}
    
    \vspace{0.4 truecm}
    \begin{tabular}{ccccccc}
    \midrule
    \multicolumn{7}{c}{JCMT SCUBA-2} \\
    \midrule
\vspace{-0.4 truecm}
UT Date & Time since & Time on & Typical & Typical & 850\,{$\mu$}m RMS & 450\,{$\mu$}m RMS  \\ \vspace{-0.4 truecm}
  & trigger & source & 225\,GHz CSO & elevation & density  &density  \\
  & (days) & (hours) & Opacity & (degrees) &(mJy/beam) & (mJy/beam) \\
\hline
2019-01-15 & 0.338 & 1.03 & 0.026 & 39 &   1.7 &    9.2 \\
2019-01-16 & 1.338 & 1.03 & 0.024 & 39 &  1.6 &    8.4 \\
2019-01-18 & 3.318 & 0.95 & 0.031 & 37 &   1.7 &   11.4 \\
\bottomrule
\end{tabular}
    \caption{{\bf Observations of GRB~190114C by ATCA and JCMT SCUBA-2}. For ATCA data, start and end date and times (UTC) of the observations, frequency, and flux (1$\sigma$ error) are reported. For JCMT SCUBA-2 data, the CSO 225 GHz opacity measures the zenith atmospheric attenuation.}
    \label{table:jcmt_data}
    \label{tab:atca}
\end{table}

\subsection{GMRT} The upgraded Giant Metre-wave Radio Telescope \cite{Swarupetal1991a} (UGMRT) observed on 17th January 2019 13.44 UT (2.8 days after the burst) in band 5 (1000-1450\,MHz) with 2048 channels spread over 400\,MHz.
GMRT detected a weak source with a flux density of 73$\pm$17 $\mu$Jy at the GRB position \cite{Cherukurieal2019a}.  
The flux should be considered as an upper limit, as the contribution from the host\cite{Tremouetal2019a} has not been subtracted.

\subsection{MeerKAT} The new MeerKAT radio observatory \cite{2018Camilo,2016jonas} observed on 15 and 18 January 2019, with DDT requested by the ThunderKAT Large Survey Project \cite{2017fender}. Both epochs used 63 antennas and were done at L-band 
spanning 856~MHz and centered at 1284~MHz. 
MeerKAT flux estimation was done by finding and fitting the source with the software PyBDSF v.1.8.15 \cite{2015mohan}.
Adding the RMS noise in quadrature to the flux uncertainty leads to final flux measurements of 125$\pm$14~$\mu$Jy/beam on 15 January and 97$\pm$16~$\mu$Jy/beam on 18 January. 
The contribution from the host galaxy\cite{Tremouetal2019a} has not been subtracted. Therefore, these measurements provide a maximum flux of the GRB.

\subsection{JCMT SCUBA-2 Sub-millimeter}
Sub-millimeter observations were performed simultaneously at 850\,{$\mu$}m and 450\,{$\mu$}m on three nights using the SCUBA-2 continuum camera\cite{Hollandetal2013}.
GRB\,190114C was not detected on any of the individual nights.
Combining all the SCUBA-2 continuum camera\cite{Hollandetal2013} observations, the RMS background noise is 0.95\,mJy/beam at 850\,{$\mu$}m and 5.4\,mJy/beam at 450\,{$\mu$}m at 1.67 days after the burst trigger.

 \begin{figure}
 \centering
 \includegraphics[width=0.9\textwidth]{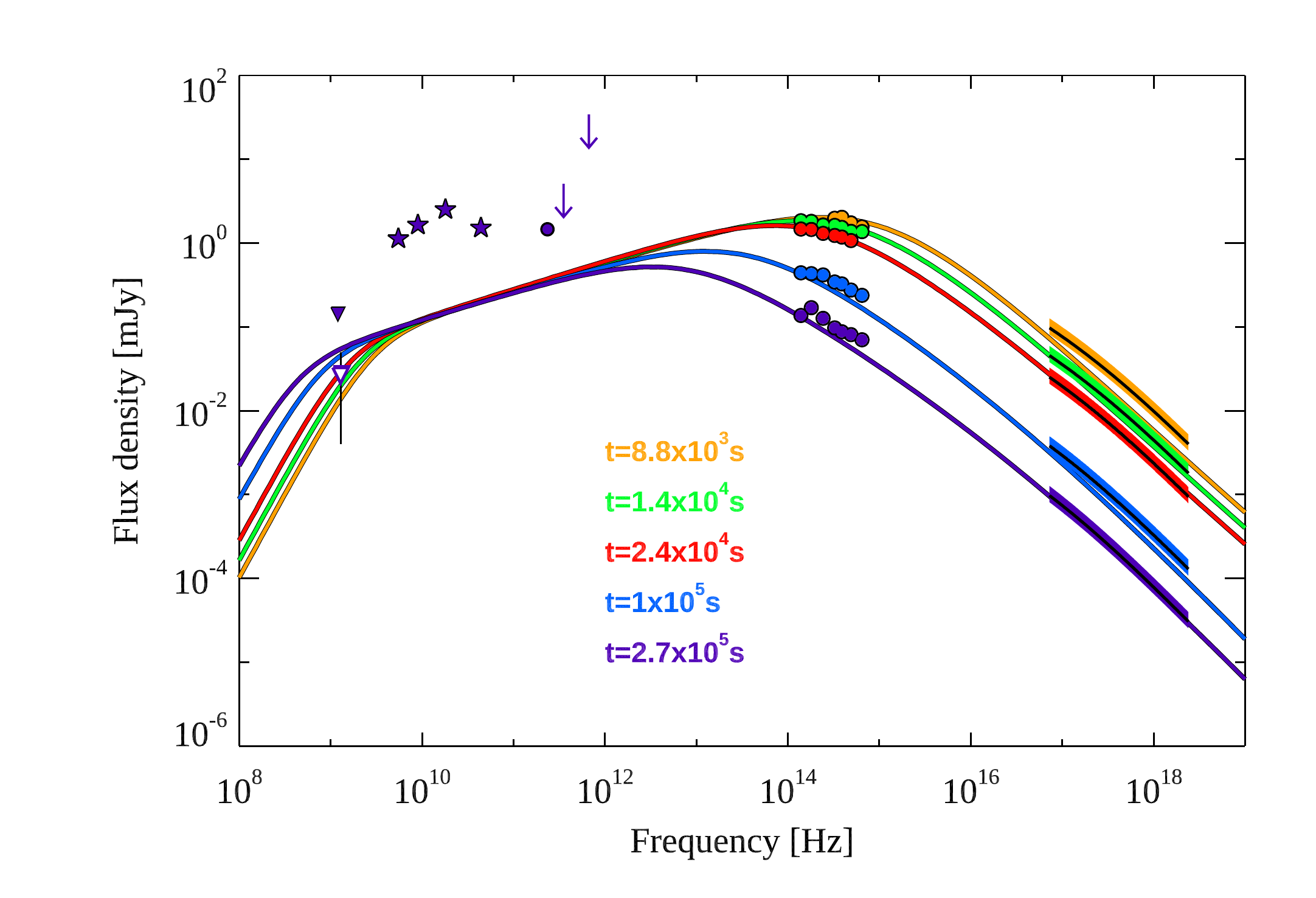}
 \caption{{\bf Radio to X-rays SED at different epochs}.
 The synchrotron frequency \nui\ crosses the optical band, moving from higher to lower frequencies. The break between $10^8$ and $10^{10}$\,Hz is caused by the self-absorption synchrotron frequency \nua. Optical (X-ray) data have been corrected for extinction (absorption).}
 \label{fig:SED_late}
 \end{figure}

\subsection{Prompt emission model for the early time MAGIC emission} 
In the standard picture the prompt sub-MeV spectrum is explained as a synchrotron radiation from relativistic accelerated electrons in the energy dissipation region. 
The associated inverse Compton component is sensitive to the details of the dynamics: e.g. in the internal shock model if the peak energy is initially very high and the IC component is suppressed due to Klein-Nishina (KN) effects, the peak of the IC component may be delayed and become bright only at late times when scatterings occur in Thomson regime. Simulations showed that magnetic fields required to produce the GeV/TeV component are rather low\cite{bosnjak2009},  $\epsilon_B \sim$10$^{-3}$. In this framework the contribution of the IC component to the observed flux at early times (62-90\,s, see Extended Data Table~\ref{tab:magicSfit}) does not exceed $\sim$ 20$\%$. Alternatively, if the prompt emission originates in reprocessed photospheric emission, the early TeV flux may arise from IC scatterings of thermal photons by freshly heated electrons below the photosphere at low optical depths. Another possibility for the generation of TeV photons might be the inverse Compton scattering of prompt MeV photons by electrons in the external forward shock region where electrons are heated to an average Lorentz factor of order 10$^4$ at early times. 

\subsection{Afterglow model}
Synchrotron and SSC radiation from electrons accelerated at the forward shock has been modelled within the external shock scenario \cite{Sarietal1998,Panaitescu&Kumar2000,Granot&Sari2002,Nakaretal2009,Sari&Esin2001}.
The results of the modeling are overlaid to the data in Fig.~\ref{fig:SED_68-180}, and Extended Data Figs.~\ref{fig:SED_late} and ~\ref{fig:lc_with_model}.

We consider two types of power law radial profiles $n(R)=n_0\,R^{-s}$ for the external environment: $s=0$ (homogeneous medium) and $s=2$ (wind-like medium, typical of an environment shaped by the stellar wind of the progenitor). In the last case, we define $n_0 =3\times 10^{35}\, A_\star$\,cm$^{-1}$.
We assume that electrons swept up by the shock are accelerated into a power law distribution described by spectral index $p$: $dN/d\gamma\propto \gamma^{-p}$, where $\gamma$ is the electron Lorentz factor.
We call $\nu_{\rm m}$ the characteristic synchrotron frequency of electrons with Lorentz factor $\gamma_{\rm m}$, \nuc\ the cooling frequency, and \nua\ the synchrotron self-absorption frequency.

The early time optical emission (up to $\sim1000$\,s) and radio emission (up to $\sim10^5$\,s) are most likely dominated by reverse shock radiation \cite{Laskaretal2019}. Detailed modeling of this component is not discussed in this work, where we focus on forward shock radiation.

\begin{figure}
    \centering
    \includegraphics[scale=0.5]{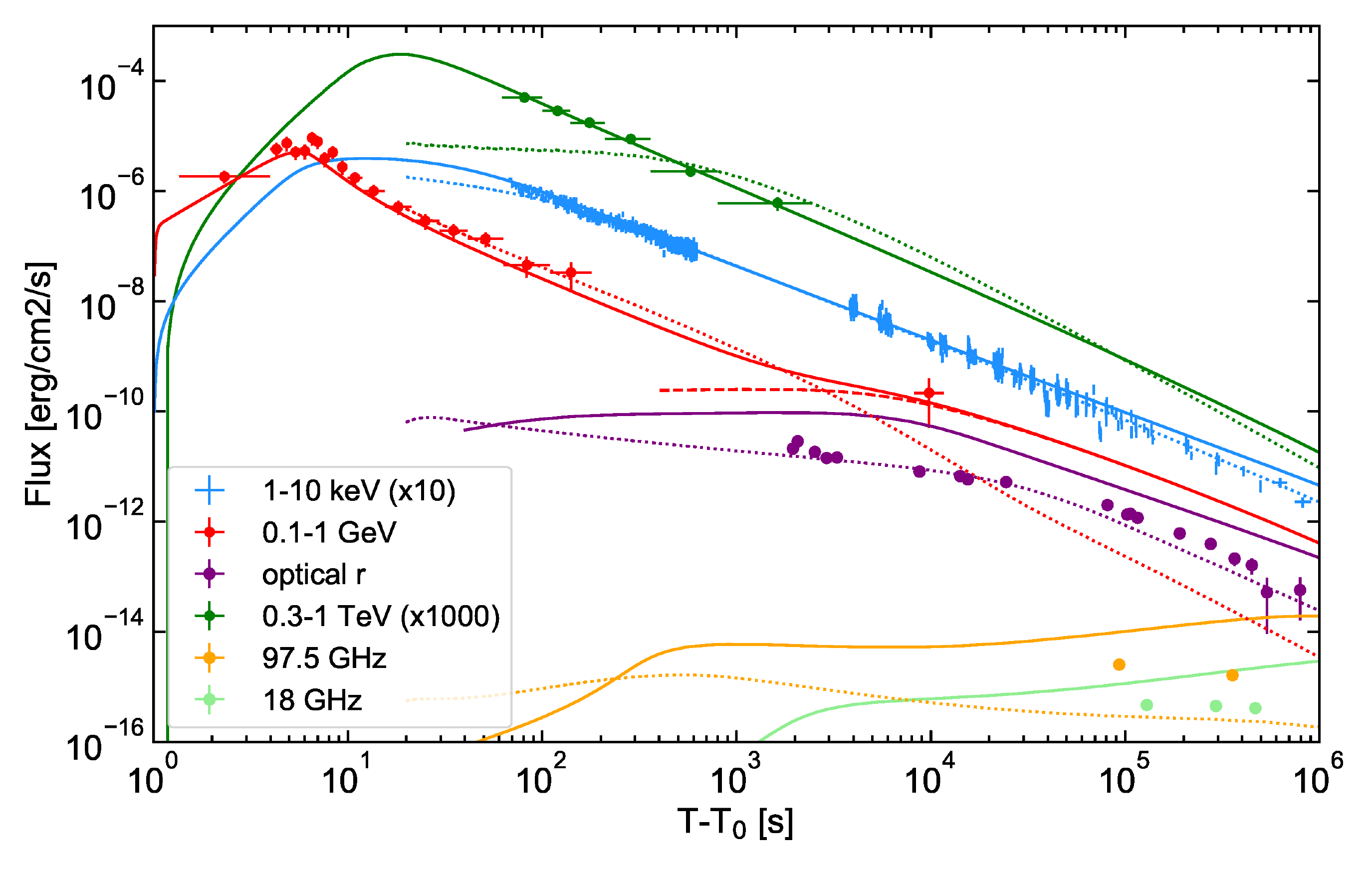}~
\caption{{\bf Modeling of the broadband light curves.} Modeling of forward shock emission is compared to observations at different frequencies (see legend). 
    The model shown with solid and dashed lines is optimised to describe the high-energy radiation (TeV, GeV and X-ray). It has been obtained with the following parameters: $s=0$, $\epsilon_{\rm e}=0.07$, $\epsilon_{\rm B}=8\times10^{-5}$, $p=2.6$, $n_0=0.5$, and $E_{\rm k}=8\times10^{53}$\,erg.
    Solid lines show the total flux (synchrotron and SSC), while the dashed line refers to the SSC contribution only. Dotted curves are derived to test a better modeling of observations at lower frequencies, but fail to explain the behaviour of the TeV light curve. These are obtained with the following model parameters: $s=2$, $\epsilon_{\rm e}=0.6$, $\epsilon_{\rm B}=10^{-4}$, $p=2.4$, $A_\star=0.1$, and $E_{\rm k}=4\times10^{53}$\,erg.}
    \label{fig:lc_with_model}
\end{figure}

The XRT flux (Fig.~\ref{fig:lc}, blue data points) decays as $F_{\rm X}\propto t^{\alpha_{\rm X}}$ with $\alpha_{\rm X}=-1.36\pm0.02$. If $\nu_{\rm X}>max(\nu_{\rm m},\nu_{\rm c})$, the X-ray light curve is predicted to decay as $t^{(2-3p)/4}$, that implies $p\sim2.5$. Another possibility is to assume $\nu_{\rm m}<\nu_{\rm X}<\nu_{\rm c}$, which implies $p=2.1-2.2$ for $s=2$ and $p\sim2.8$ for $s=0$.
A broken power law fit provides a better fit ($5.3\times10^{-5}$ probability of chance improvement), with a break occurring around $4\times10^4$\,s and decay indices ${\alpha_{\rm X,1}}\sim-1.32\pm0.03$ and ${\alpha_{\rm X,2}}\sim-1.55\pm0.04$.
This behaviour can be explained by the passage of $\nu_{\rm c}$ in the XRT band and assuming again $p=2.4-2.5$ for $s=2$ and $p\sim2.8$ for $s=0$. 

The optical light curve starts displaying a shallow decay in time (with temporal index poorly constrained, between -0.5 and -0.25) starting from $\sim 2\times10^3$\,s, followed by a steepening around $8\times10^4$\,s, when the temporal decay becomes similar to the decay in X-ray band, suggesting that after this time the X-ray and optical band lie in the same part of the synchrotron spectrum. If the break is interpreted as the synchrotron characteristic frequency \nui\ crossing the optical band, after the break the observed temporal decay requires a steep value of $p\sim3$ for $s=0$ and a value between $p=2.4$ and $p=2.5$ for $s=2$.
Independently of the density profile of the external medium and on the cooling regime of the electrons, $\nu_{\rm m}\propto t^{-3/2}$, placing it the soft X-ray band at $10^2$\,s.
The SED at $\sim$100\,s is indeed characterised by a peak in between 5-30\,keV (Fig.~\ref{fig:SED_68-180}).
Information on the location of the self-absorption frequency are provided by observations at 1\,GHz, showing that \nua$\sim 1$\,GHz at $10^5$\,s (Extended Data Fig.~\ref{fig:SED_late}).

Summarizing, in a wind-like scenario X-ray and optical emission and their evolution in time can be explained if $p=2.4 - 2.5$, the emission is initially in fast cooling regime and transitions to a slow cooling regime around $3\times10^3$\,s.
The optical spectral index at late times is predicted to be $(1-p)/2\sim-0.72$, in agreement with observations.
\nui\ crosses the optical band at $t\sim8\times10^4$\,s, explaining the steepening of the optical light curve and the flattening of the optical spectrum. The X-ray band initially lies above (or close to) $\nu_{\rm m}$, and the break frequency \nuc\ starts crossing the X-ray band around $2-4\times10^4$\,s, producing the steepening in the decay rate (the cooling frequency increases with time for $s=2$).
In this case, before the temporal break, the decay rate is related to the spectral index of the electron energy distribution by $\alpha_{\rm X,1}=(2-3p)/4\sim-1.3$, for $p\sim2.4-2.5$. Well after the break, this value of $p$ predicts a decay rate $\alpha_{\rm X,1}=(1-3p)/4=-1.55 - 1.62$.
Overall, this interpretation is also consistent with the fact that the late time ($t>10^5$\,s) X-ray and optical light curves display similar temporal decays (Fig.\ref{fig:lc}), as they lie in the same part of the synchrotron spectrum ($\nu_{\rm m}<\nu_{\rm opt}<\nu_{\rm X}<\nu_{\rm c}$).
A similar picture can be invoked to explain the emission also assuming a homogeneous density medium, but a steeper value of $p$ is required.
In this case, however, no break is predicted in the X-ray light curve.

We now add to the picture the information brought by the TeV detection.
The modeling is built with reference to the MAGIC flux and spectral indices derived considering statistical errors only (see Extended Data Table~\ref{tab:magicSfit} and green data points in Extended Data Fig.~\ref{fig:sederr}).
The light curve decays in time as $t^{-1.51}$ and the photon index is consistent within $\sim1\sigma$ with $\Gamma_{\rm ph, TeV}\sim -2.5$ for the entire duration of the emission, although there is evidence for an evolution from harder ($\sim-2$) to softer ($\sim-2.8$) values.
In the first broadband SED (Fig.~\ref{fig:SED_68-180}, 68-110\,s), LAT observations provide strong evidence for the presence of two separated spectral peaks. 

Assuming Thomson scattering, the SSC peak is given by:
\begin{equation}
    {\nu^{\rm SSC}_{\rm peak}}\simeq2\,\gamma^2_{\rm e}{\nu^{\rm syn}_{\rm peak}}\,
\end{equation}
while in KN regime, the SSC peak should be located at:
\begin{equation}
     h\nu^{\rm SSC}_{\rm peak} \simeq 2\,\gamma_{\rm e}\,\Gamma\,m_{\rm e}\,c^2/ (1+z)\,
\end{equation}
where $\gamma_{\rm e}=\min(\gamma_{\rm c},\gamma_{\rm m})$.
The synchrotron spectral peak is located at $E^{\rm syn}_{\rm peak}\sim10$\,keV, while the peak of the SSC component must be below $E^{\rm ssc}_{\rm peak}\lesssim100$\,GeV to explain the MAGIC photon index. Both the KN and Thomson scattering regimes imply $\gamma_{\rm e}\lesssim10^3$.
This small value faces two problems:
i) if the bulk Lorentz factor $\Gamma$ is larger than 150 (that is a necessary condition to avoid strong \pp\ opacity, see below), a small \gm\ translates into a small efficiency of the electron acceleration, with $\epsilon_{\rm e}<0.05$, ii) the synchrotron peak energy can be located at $E^{\rm syn}_{\rm peak}\sim10$\,keV only for $B\,\Gamma \gtrsim 10^5$\,G.
Large $B$ and small \ee\ would make difficult to explain the presence of a strong SSC emission.
These calculations show that \pp\ opacity likely plays a role in shaping and softening the observed spectra of the SSC spectrum\cite{DerishevPiran2019,WangZhang2019}. 

For a gamma-ray photon with energy $E_{\gamma}$, the $\tau_{\gamma\gamma}$ opacity is:
\begin{equation}
    \tau_{\gamma\gamma}(E_{\gamma})=\sigma_{\gamma\gamma}\,(R/\Gamma)\,n_{\rm t}(E_\gamma)\,,
\end{equation}
where $n_{\rm t}=L_{\rm t}/(4\,\pi\,R^2\,c\,\Gamma\,E_{\rm t})$ is the density of target photons in the comoving frame, $L_{\rm t}$ is the luminosity and $E_{\rm t}=(m_{\rm e}\,c^2)^2\,\Gamma^2/E_{\gamma}/(1+z)^2$ is the energy of target photons in the observer frame.
Target photons for photons with energy $E_\gamma=0.2-1$\,TeV  and for $\Gamma\sim120-150$ have energies in the range $4-30$\,keV.
When $\gamma-\gamma$ absorption is relevant, the emission from pairs can give a non-negligible contribution to the radiative output.

To properly model all the physical processes that are shaping the broadband radiation, we use a numerical code that solves the evolution of the electron distributions and derives the radiative output taking into account the following processes: synchrotron and SSC losses, adiabatic losses, $\gamma-\gamma$ absorption, emission from pairs, and synchrotron self-absorption\cite{MastichiadisKirk1995,VurmPoutanen2009,PetropoulouMastichiadis2009,Pennanenetal14}.
We find that for the parameters assumed in the proposed modeling (see below), the contribution from pairs to the emission is negligible.

The MAGIC photon index (Extended Data Table~\ref{tab:magicSfit}) and its evolution with time constrain the SSC peak energy to be at $\lesssim1$\,TeV at the beginning of observations (Extended Data Table~\ref{tab:magicSfit}). In general the internal opacity decreases with time and KN effects become less relevant. A possible softening of the spectrum with time, as the one suggested by the observations, requires that the spectral peak decreases with time and moves below the MAGIC energy range.  
In the slow cooling regime, the SSC peak evolves to higher frequencies for a wind-like medium and decreases very slowly ($\nu^{\rm SSC}_{\rm peak}\propto t^{-1/4}$) for a constant-density medium (both in KN and Thomson regimes). In fast cooling regime the evolution is faster ($\nu^{\rm SSC}_{\rm peak}\propto t^{-1/2} - t^{-9/4}$ depending on medium and regime). 

We model the multi-band observations considering both $s=0$ and $s=2$. 
The results are shown in Fig.~\ref{fig:SED_68-180}, Extended Data Figs.~\ref{fig:SED_late} and \ref{fig:lc_with_model} where model curves are overlaid to observations. 
The model curves shown in these figures have been derived using the following parameters.
The model in Fig.~\ref{fig:SED_68-180} and in \ref{fig:lc_with_model} (solid and dashed curves) we have used $s=0$, $\epsilon_{\rm e}=0.07$, $\epsilon_{\rm B}=8\times10^{-5}$, $p=2.6$, $n_0=0.5$, and $E_{\rm k}=8\times10^{53}$\,erg. For the dotted curves in Extended Data Fig.~\ref{fig:lc_with_model} and the SEDs in Extended Data Figs.~\ref{fig:SED_late} we have used $s=2$, $\epsilon_{\rm e}=0.6$, $\epsilon_{\rm B}=10^{-4}$, $p=2.4$, $A_\star=0.1$, and $E_{\rm k}=4\times10^{53}$\,erg.

Using the constraints on the afterglow onset time ($t_{\rm peak}^{\rm aft}\sim 5-10$\,s from the smooth component detected during the prompt emission) the initial bulk Lorentz factor is constrained to assume values $\Gamma_0\sim300$ and $\Gamma_0\sim700$ for $s=2$ and $s=0$, respectively.

Consistently with the qualitative description above, we find that late time optical observations can indeed be explained with \nui\ crossing the band (see the SED modeling in Extended Data Fig.~\ref{fig:SED_late} and dotted curves in Extended Data Fig.~\ref{fig:lc_with_model}). However a large \nui\ is required in this case and consequently also the peak of the SSC component would be large and lie above the MAGIC energy range. The resulting MAGIC light curve (green dotted curve in Extended Data Fig.~\ref{fig:lc_with_model}) does not agree with observations. Relaxing the requirement on \nui, the TeV spectra (Fig.~\ref{fig:SED_68-180}) and light curve (green solid curve in Extended Data Fig.~\ref{fig:lc_with_model}) can be explained. As noted before, a wind-like medium can explain the steepening of the X-ray light curve at $8\times10^4$\,s, while in a homogeneous medium no steepening is expected (blue dotted and solid lines in Extended Data Fig.~\ref{fig:lc_with_model}). We find that the GeV flux detected by LAT at late time ($t\sim10^4$\,s) is dominated by the SSC component (dashed line in Extended Data Fig.~\ref{fig:lc_with_model}).\\
\end{methods}

\bibliography{MWL}

\begin{addendum}

\item[Author Information] The authors declare no competing interests. Correspondence and requests for materials should be addressed to Razmik Mirzoyan (email: razmik.mirzoyan@mpp.mpg.de) or MAGIC (email: contact.magic@mpp.mpg.de).

 \item The MAGIC Collaboration would like to thank the Instituto de Astrof\'{\i}sica de Canarias for the excellent working conditions at the Observatorio del Roque de los Muchachos in La Palma. The financial support of the German BMBF and MPG, the Italian INFN and INAF, the Swiss National Fund SNF, the ERDF under the Spanish MINECO (FPA2017-87859-P, FPA2017-85668-P, FPA2017-82729-C6-2-R, FPA2017-82729-C6-6-R, FPA2017-82729-C6-5-R, AYA2015-71042-P, AYA2016-76012-C3-1-P, ESP2017-87055-C2-2-P, FPA2017‐90566‐REDC), the Indian Department of Atomic Energy, the Japanese JSPS and MEXT, the Bulgarian Ministry of Education and Science, National RI Roadmap Project DO1-153/28.08.2018 and the Academy of Finland grant nr. 320045 is gratefully acknowledged. This work was also supported by the Spanish Centro de Excelencia ``Severo Ochoa'' SEV-2016-0588 and SEV-2015-0548, and Unidad de Excelencia ``Mar\'{\i}a de Maeztu'' MDM-2014-0369, by the Croatian Science Foundation (HrZZ) Project IP-2016-06-9782 and the University of Rijeka Project 13.12.1.3.02, by the DFG Collaborative Research Centers SFB823/C4 and SFB876/C3, the Polish National Research Centre grant UMO-2016/22/M/ST9/00382 and by the Brazilian MCTIC, CNPq and FAPERJ.
L. Nava acknowledges funding from the European Union's Horizon 2020 Research and Innovation programme under the Marie Sk\l odowska-Curie grant agreement n.\,664931.
E. Moretti acknowledges funding from the European Union's Horizon 2020 research and innovation programme under Marie Sk\l odowska-Curie grant agreement No 665919.
 This paper makes use of the following ALMA data:
 ADS/JAO.ALMA\#2018.A.00020.T, 
 \linebreak ADS/JAO.ALMA\#2018.1.01410.T. ALMA is a partnership of ESO (representing its member states), NSF (USA) and NINS (Japan), together with NRC (Canada), MOST and ASIAA (Taiwan), and KASI (Republic of Korea), in cooperation with the Republic of Chile. The Joint ALMA Observatory is operated by ESO, AUI/NRAO and NAOJ.
 CT, AdUP, and DAK acknowledge support from the Spanish research project AYA2017-89384-P. C. Thoene and A. de Ugarte Postigo acknowledge support from funding associated to Ram\'on y Cajal fellowships (RyC-2012-09984 and RyC-2012-09975). D. A. Kann acknowledges support from funding associated to Juan de la Cierva Incorporaci\'on fellowships (IJCI-2015-26153).
 The James Clerk Maxwell Telescope is operated by the East Asian Observatory on behalf of The National Astronomical Observatory of Japan; Academia Sinica Institute of Astronomy and Astrophysics; the Korea Astronomy and Space Science Institute; Center for Astronomical Mega-Science (as well as the National Key R\&D Program of China with No. 2017YFA0402700). Additional funding support is provided by the Science and Technology Facilities Council of the United Kingdom and participating universities in the United Kingdom and Canada. The JCMT data reported here were obtained under project M18BP040 (P.I. D. Perley). We thank Mark Rawlings, Kevin Silva, Sheona Urquart, and the JCMT staff for the prompt support of these observations.
 The Liverpool Telescope, located on the island of La Palma in the Spanish Observatorio del Roque de los Muchachos of the Instituto de Astrofisica de Canarias, is operated by Liverpool John Moores University with financial support from the UK Science and Technology Facilities Council. 
 The Australia Telescope Compact Array is part of the Australia Telescope National Facility which is funded by the Australian Government for operation as a National Facility managed by CSIRO. GEA is the recipient of an Australian Research Council Discovery Early Career Researcher Award (project number DE180100346) and JCAM-J is the recipient of Australian Research Council Future Fellowship (project number FT140101082) funded by the Australian Government.  
Support for the German contribution to GBM was provided by the Bundesministerium f{\"u}r Bildung und Forschung (BMBF) via the Deutsches Zentrum f{\"u}r Luft und Raumfahrt (DLR) under grant number 50 QV 0301. The UAH coauthors gratefully acknowledge NASA funding from cooperative agreement NNM11AA01A. C.A.W.H., and C.M.H. gratefully acknowledge NASA funding through the {\it Fermi}-GBM project.  

The \textit{Fermi} LAT Collaboration acknowledges generous ongoing support from a number of agencies and institutes that have supported both the development and the operation of the LAT as well as scientific data analysis. These include the National Aeronautics and Space Administration and the
Department of Energy in the United States, the Commissariat \`a l'Energie Atomique and the Centre National de la Recherche Scientifique / Institut National de Physique
Nucl\'eaire et de Physique des Particules in France, the Agenzia Spaziale Italiana and the Istituto Nazionale di Fisica Nucleare in Italy, the Ministry of Education,
Culture, Sports, Science and Technology (MEXT), High Energy Accelerator Research Organization (KEK) and Japan Aerospace Exploration Agency (JAXA) in Japan, and
the K.~A.~Wallenberg Foundation, the Swedish Research Council and the Swedish National Space Board in Sweden.
 
Additional support for science analysis during the operations phase is gratefully
acknowledged from the Istituto Nazionale di Astrofisica in Italy and the Centre
National d'\'Etudes Spatiales in France. This work performed in part under DOE
Contract DE-AC02-76SF00515.

Part of the funding for GROND (both hardware as well as personnel) was generously granted from the Leibniz-Prize to Prof. G. Hasinger (DFG grant HA 1850/28-1). 
 Swift data were retrieved from the Swift archive at HEASARC/NASA-GSFC, and from the UK Swift Science Data Centre. Support for Swift in the UK is provided by the UK Space Agency %
 
 This work is based on observations obtained with XMM-Newton, an ESA science mission with instruments and contributions directly funded by ESA Member States and NASA. 

This work is partially based on observations collected at the European Organisation for Astronomical Research in the Southern Hemisphere under ESO programme 199.D-­‐‑0143. 
The work is partly based on observations made with the Gran Telescopio Canarias (GTC), installed in the Spanish Observatorio del Roque de los Muchachos of the Instituto de Astrof\'isica de Canarias, in the island of La Palma.
This work is partially based on observations made with the Nordic Optical Telescope (programme 58-502), operated by the Nordic Optical Telescope Scientific Association at the Observatorio del Roque de los Muchachos, La Palma, Spain, of the Instituto de Astrof\'isica de Canarias.
This work is partially based on observations collected at the European Organisation for Astronomical Research in the Southern Hemisphere under ESO programme 102.D-0662.
This work is partially based on observations collected through the ESO programme 199.D-0143 ePESSTO.
M. Gromadzki is supported by the Polish NCN MAESTRO grant 2014/14/A/ST9/00121.
M. Nicholl is supported by a Royal Astronomical Society Research Fellowship
M.~G. Bernardini, S. Campana, A. Melandri and P. D'Avanzo acknowledge ASI grant I/004/11/3.
S. Campana thanks for support the implementing agreement ASI-INAF n.2017-14-H.0. S.~J.~Smartt acknowledges funding from STFC Grant Ref: ST/P000312/1. 
NPMK acknowledges support by the UK Space Agency under grant ST/P002323/1 and the UK Science and Technology Facilities Council under grant ST/N00811/1.
L. Piro, S. Lotti acknowledge partial support from the agreement ASI-INAF n.2017-14-H.0.
VAF acknowledges RFBR 18-29-21030 for support.
AJCT acknowledges support from the Junta de Andaluc\'ia (Project P07-TIC-03094) and support from the Spanish Ministry Projects AYA2012-39727-C03-01 and 2015-71718R.
KM acknowledges the support from Department of Science and Technology (DST), Govt. of India and Indo-US Science and Technology Forum (IUSSTF) for the WISTEMM fellowship and Dept. of Physics, UC Davis where a part of this work was carried out.  M.J.M.~acknowledges the support of the National Science Centre, Poland through the grant 2018/30/E/ST9/00208. VJ and RL acknowledges support from the grant EMR/2016/007127 from Dept. of Science and Technology, India.
K.~Maguire acknowledges support from H2020 through an ERC Starting Grant (758638).
L. Izzo would like to acknowledge Massimo Della Valle for invaluable support in the operation of the telescope.

 \item[Author Contributions] The MAGIC telescope system was designed and constructed by the MAGIC Collaboration.  Operation, data processing, calibration, Monte Carlo simulations of the detector, and of theoretical models, and data analyses were performed by the members of the MAGIC Collaboration, who also discussed and approved the scientific results. 
L.\,Nava coordinated the gathering of the data, developed the theoretical interpretation, and wrote the main section and the section on Afterglow Modeling.
E.\,Moretti coordinated the analysis of the MAGIC data, wrote the relevant sections, and, together with F.\,Longo, coordinated the collaboration with the Fermi team.
D.\,Miceli, Y.\,Suda and S.\,Fukami performed the analysis of the MAGIC data. 
S.\,Covino provided support with the analysis of the optical data and the writing of the corresponding sections.
Z.\,Bosnjak performed calculations for the contribution from prompt emission to TeV radiation and wrote the corresponding section.
A.\,Stamerra, D.\,Paneque and S.\,Inoue contributed in structuring and editing the paper. 
A.\,Berti contributed to editing and finalising the manuscript.
R.\,Mirzoyan coordinated and supervised the paper.
All MAGIC collaborators contributed to the editing and comments to the final version of the manuscript.

 S.~Campana and M.~G.~Bernardini extracted the spectra and performed the spectral analysis of {\it Swift}/BAT and {\it Swift}/XRT data.
 N.~P.~M.~Kuin derived the photometry for the {\it Swift}/UVOT event mode data, and the uv grism exposure.
 M.~H.~Siegel derived the image mode Swift UVOT photometry.
  A. de Ugarte Postigo was principal investigator of ALMA program 2018.1A.00020.T, triggered these observations and performed photometry. S. Martin reduced the ALMA Band 6 data. C. C. Th\"one, S. Schulze, D. A. Kann, and M. Micha{\l}owski participated in the ALMA DDT proposal preparation, observations, and scientific analysis of the data.
  D. A. Perley was principal investigator of ALMA program 2018.1.01410.T and triggered these observations, and was also principal investigator of the LT programme and the JCMT programme.  A. M. Cockeram analyzed the ALMA Band 3 and LT data, and wrote the LT text.
 S. Schulze contributed to the development of the ALMA Band 3 observing programme. I. A. Smith triggered the JCMT programme, analyzed the data, and wrote the associated text.  N. R. Tanvir contributed to the development of the JCMT programme. D. A. Kann and C. C. Th\"one triggered and coordinated the X-shooter 
observations. D. A. Kann independently checked the optical light curve 
analysis.
  K. Misra was the principal investigator of the GMRT programme 35\_018. S. V. Cherukuri and V. Jaiswal analyzed the data. L. Resmi contributed to the observation plan and data analysis.
 E.T., I.H. and R.D. have performed the MeerKAT data analysis. 
 G. Anderson, A. Moin, S. Schulze and E. Troja were principal investigator of ATCA program CX424. G. Anderson, M. Wieringa and J. Stevens carried out the observations. G. Anderson, G. Bernardi, S. Klose, M. Marongiu, A. Moin, R. Ricci and M. Wieringa analysed these data. M. Bell, J. Miller-Jones and L. Piro participated to the ATCA proposal preparation and scientific analysis of the data.
 The ePESSTO project was delivered by the following who have contributed to managing, executing, reducing, analysing ESO/NTT data and provided comments to the manuscript: J.~P.~Anderson, N.~Castro Segura, P.~D'Avanzo, M.~Gromadzki, C.~Inserra, E.~Kankare, K.~Maguire, M.~Nicholl, F.~Ragosta, S.~J.~Smartt. 
A.~Melandri and A.~Rossi reduced and analyzed REM data and provided comments to the manuscript.  
J.~Bolmer was responsible for observing the GRB with GROND as well as for the data reduction and calibration. J.~Bolmer and J.~Greiner contributed to the analysis of the data and writing of the text.
 E. Troja triggered the {\it NuSTAR} TOO observations performed under DDT program, L. Piro requested the XMM-Newton data carried out under DDT program and carried out the scientific analysis of XMM-Newton and {\it NuSTAR}. S. Lotti analyzed the {\it NuSTAR} data and wrote the associated text. A. Tiengo and G. Novara analysed the XMM-Newton data and wrote the associated text.
AJCT led the observing BOOTES and GTC programs. AC, CJPP, EFG, IMC, SBP and XYL analyzed the BOOTES data whereas AFV, MDCG, RSR, YDH and VVS analyzed the GTC data and interpreted them accordingly.
N.~Tanvir created the X-shooter and AlFOSC figures.
J.~Fynbo, J. Japelj performed the analysis of X-shooter and AlFOSC spectra.
D.~Xu, P.\,Jakobsson contributed to NOT programme and triggering.
D.~Malesani performed photometric analysis of NOT data.
E.\,Peretti contributed to developing the code for modeling of afterglow radiation. 
L. Izzo triggered and analysed the OASDG data, while A. Di Dato and A. Noschese executed the observations at the telescope.

\item[Data Availability Statement] Data are available from the corresponding authors upon request.

\item[Code Availability Statement] Proprietary data reconstruction codes were generated at the MAGIC telescopes large-scale facility. Information supporting the findings of this study are available from the corresponding authors upon request.

\end{addendum}

{%
\let\oldnumberline\numberline%
\renewcommand{\numberline}{\figurename~\oldnumberline}%
}

\end{document}